\documentclass[a4paper,11pt]{article}
\pdfoutput=1 

\usepackage{jheppub} 

\usepackage{graphicx}
\usepackage{epstopdf}
\usepackage{amsmath}
\usepackage{amsxtra}
\usepackage{amstext}
\usepackage{amssymb}
\usepackage{subfigure}
\usepackage{slashed}
\usepackage{latexsym} 
\usepackage{pdfpages}
\usepackage{hyperref}

\title{Efficient Monte Carlo Integration Using Boosted Decision Trees and Generative Deep Neural Networks}
\author{Joshua Bendavid}
\affiliation{Lauritsen Laboratory for High Energy Physics, California Institute of Technology}
 
\emailAdd{Josh.Bendavid@cern.ch}
 
\abstract{New machine learning based algorithms have been developed and tested for Monte Carlo integration based on generative Boosted Decision Trees and Deep Neural Networks.  Both of these algorithms exhibit substantial improvements compared to existing algorithms for non-factorizable integrands in terms of the achievable integration precision for a given number of target function evaluations.  Large scale Monte Carlo generation of complex collider physics processes with improved efficiency can be achieved by implementing these algorithms into commonly used matrix element Monte Carlo generators once their robustness is demonstrated and performance validated for the relevant classes of matrix elements.}

\begin{document}
\maketitle
\flushbottom

\section{Introduction}
The problem of evaluating the integral of an arbitrary function over a specified domain is common in high energy physics.  A frequent use case is the calculation of a total interaction cross section given a matrix element which is computable point-wise in the phase-space of the incoming and outgoing particle momenta.  For cases with many outgoing particles, such as the simulation of processes with several additional jets at hadron colliders, the dimensionality of this integral can be large.  Moreover, when the matrix element includes next to leading order contributions or beyond, the number of sub-contributions are large, and may involve numerical integrals for the evaluation of loops, and therefore the point-wise evaluation of the matrix element can be computationally intensive.  This motivates the development of algorithms for multidimensional numerical integration which are efficient, in terms of maximizing the precision of the result for the smallest possible number of function evaluations.

Existing algorithms are typically variations of Monte Carlo integration with importance sampling.  The most common are variations on VEGAS\cite{vegas} which is limited in efficiency by the degree to which the integrand can be factorized into a product of one-dimensional distributions.  This limitation can be mitigated by some combination of transformation of the integration space to better factorize the integrand, and the use of multi-channeling techniques to decompose the integrand into a sum of (more) factorizable contributions\cite{vamp}.  Algorithms which improve the underlying efficiency for difficult to factorize integrands have been developed in the form of FOAM\cite{foam}, based on dividing the phase space into hypercubes, each sampled uniformly with an appropriate weight.  This algorithm has a close correspondence to binary decision trees as used for classification or regression problems.  For this class of applications, it is well known that the extension of single decision tree to an ensemble of decision trees via boosting methods can greatly improve the performance\cite{Friedman98additivelogistic,Friedman00greedyfunction}.  We develop an algorithm loosely based on the FOAM principle, but extending the partitioning of the phase space from a single binary decision tree, to a boosted ensemble in order to improve the efficiency of the integration, analogous to the improvement in performance for the classification and regression case.  In addition to the Boosted Decision Tree (BDT) based algorithm, we develop an orthogonal approach based on generative deep neural networks, which can be considered a multi-dimensional and non-analytic generalization of inverse transform sampling techniques.

\section{Gradient Boosted Regression Integration}
Importance-sampling-based integration algorithms require a probability distribution which is easily sampled from, where the probability density associated with each sampled $d$-dimensional point $\bar x$ can be easily evaluated, and which well-approximates the probability density associated with the function to be integrated $f(\bar x)$.  In FOAM and similar algorithms, this is achieved by constructing a single binary decision tree with an output value given by
\begin{equation}
 g(\bar x) = \sum_{j} \theta_j(\bar x) a_j
\end{equation}

such that $g(\bar x)\approx f(\bar x)$, where $\theta_j$ are a set of non-overlapping multidimensional box functions, each covering a $d$-dimensional hyper-cubic region of phase space, defined by the set of decision tree splits, and $a_j$ are a set of weights.  In this case the normalized probability density associated with $g(\bar x)$ is given by the normalized sum over hyper-cubes
\begin{equation}
 p_g(\bar x) = \frac{1}{\sum_i V_j a_j}\sum_{j} \theta_j(\bar x) a_j
\end{equation}
where $V_j$ is the volume of each hypercube.  This density can be easily sampled by first randomly choosing a hypercube with probability proportional to $V_j a_j$, and then sampling uniformly over the corresponding hypercube volume.

This can be extended to an additive series of decision trees $g_i$ by defining the output function $g(\bar x)$.
\begin{align}
 g(\bar x) &= \sum_i g_i(\bar x)\\
	   &= \sum_i \sum_j \theta_{ij}(\bar x) a_{ij}
\end{align}

In this case the corresponding probability density is
\begin{align}
 p_g(\bar x) = \frac{1}{\sum_i \sum_j V_{ij} a_{ij}}\sum_i \sum_j \theta_{ij}(\bar x) a_{ij}
\end{align}
This probability density can again be efficiently sampled from by first randomly choosing a hypercube with probability proportional to $V_{ij} a_{ij}$ and then sampling uniformly over the corresponding hypercube volume, where the difference with respect to the single-tree case is that the hypercubes are chosen from amongst all the trees in the series, and therefore may overlap in general.  This sampling procedure imposes an important constraint with respect to the general case, where efficient sampling requires that all of the weights $a_{ij}$ are non-negative\footnote{Sampling from the decision tree series is still possible in the case of negative weights, but this requires sampling independently from positive and negative-weighted hypercubes and introduces undesirable statistical properties.}.

Constructing a decision tree series of this kind optimized for numerical integration of $f(\bar x)$ requires defining a loss function to be minimized by gradient boosting as in \cite{Friedman00greedyfunction}.  The integration weight for each sampled phase space point can be defined as the ratio $w = f(\bar x)/p_g(\bar x)$.  In this case the precision of the integral $I_f = \int f(\bar x)d\bar x$ for $N$ sampled points is given by 
\begin{equation}
 \sigma(I_f) = \frac{1}{\sqrt{N}}\frac{1}{<w>}\left<\left(w - <w>\right)^2\right>
 \label{eqn:uncert}
\end{equation}
where the expectation values are taken over the sampled phase space points.  While minimizing this expression with respect to the decision tree boundaries and weights for fixed $N$ would yield the optimal integration precision, this is not practical in the gradient boosting context, given that it cannot be expressed as a sum of independent contributions for each phase-space point.  A more convenient loss function which approximates the above can be defined
\begin{equation}
 L = \sum\left(\ln f(\bar x) - \ln g(\bar x)\right)^2
\end{equation}
which corresponds to fitting the mean of a log-normal distribution.  While this loss function can be expressed as a sum of independent contributions, it is still difficult to minimize by gradient boosting given the presence of non-convex regions which preclude using second-order methods for the minimization.  A further modified loss function is therefore defined at each $i^{th}$ iteration of the gradient boosting procedure
\begin{equation}
 L_i = \sum\left[\ln\left(\textnormal{max}\left(f(\bar x) - \sum_{k=0}^{i-1}t_k(\bar x),\epsilon\right)\right) - \ln g_i(\bar x)\right]^2
\end{equation}
which corresponds to fitting the mean of a log-normal distribution with respect to the \textit{residuals} after the previous iterations, where $\epsilon$ is a numerical cutoff to make the computation well-defined.  This has the advantage that $g(\bar x)$ constructed in this way is guaranteed to have non-negative weights, as required for easy sampling.

Since negative weights are not possible, there is no mechanism for later trees in the series to compensate for a series of earlier trees which may overshoot the target function.  For this reason, very slow convergence is required to achieve a well optimized set of trees.  For this reason, an additional BDT $h(\bar x)$ is introduced such that $e^{h(\bar x)} \approx f(\bar x)$.  This BDT can be constructed according to the standard training procedure for regression, with the loss function $L_h$ as below, which recovers the desired log-normal constraint for $e^{h(\bar x)}$ expressed in terms of a simple Gaussian constraint on $h(\bar x)$ which is much simpler to minimize by gradient boosting.
\begin{equation}
 L_h = \sum\left(\ln f(\bar x) - h(\bar x)\right)^2
 \label{eqn:lognorm}
\end{equation}
In this case, the two BDT's can be trained in parallel, with $e^{h(\bar x)}$ replacing $f(\bar x)$ in the loss function for the non-negative BDT such that
\begin{equation}
 L_i \rightarrow \sum\left[\ln\left(\textnormal{max}\left(e^{h(\bar x)} - \sum_{k=0}^{i-1}t_k(\bar x),\epsilon\right)\right) - \ln g_i(\bar x)\right]^2
\end{equation}

In order to construct an optimized set of BDT's while minimizing the number of required function evaluations of $f(\bar x)$, an iterative procedure is defined as follows:
\begin{enumerate}
 \item Create an initial training dataset by sampling from an easily sampled prior distribution (by default a uniform distribution over the specified integration integration range) and evaluating the target function value $f(\bar x)$ for each sampled point.
 \item Iteratively sample and train BDT's:
 \begin{enumerate}
 \item Iterative train BDT's for $N$ trees
 \begin{enumerate}
  \item Train one additional tree for standard regression BDT $h(\bar x)$
  \item Train one additional tree for non-negative generative BDT $g(\bar x)$
  \end{enumerate}
  \item Check for convergence and end training if appropriate (for example checking the relative change in BDT integral)
  \item Add to training dataset by sampling additional events from the generative BDT $g(\bar x)$ and evaluating target function $f(\bar x)$ for each sampled point
 \end{enumerate}
\end{enumerate}

Once the training of the BDTs is complete, numerical integration of the target function can be performed by sampling an arbitrary number of phase-space points from the generative BDT $g(\bar x)$ until the desired precision is reached, where as usual for importance-sampling, the value of the integral is given by the mean of the integration weights $<f(\bar x)/p_g(\bar x)>$ and the uncertainty is given by (\ref{eqn:uncert}).

Since in general the training of the standard regression BDT is easier, and the resulting precision better as compared to the generative BDT, the integration can also instead be carried out using a staged approach, where the generative BDT is used to evaluate the integral of the regression approximation $e^{h(\bar x)}$ as well as sample from it using accept-reject sampling.  These samples can the be used to evaluate the integral of $f(\bar x)$ with a better precision for a fixed number of function evaluations, at the expense of additional sampling from $g(\bar x)$ and additional evaluations of the BDT values $g(\bar x)$ and $h(\bar x)$, which might nevertheless be worthwhile in the case where evaluating $f(\bar x)$ is computationally intensive.

\subsection{Results}

Performance tests are carried out using the Camel function as in\cite{vegas}, consisting of a pair of $d$-dimensional gaussians with $\sigma = \frac{1}{10\sqrt{2}}$ and means placed at $1/3$ and $2/3$ along the multidimensional diagonal of a unit-hypercube integration region.

\begin{figure}[htb!]
 \includegraphics[width=0.5\textwidth]{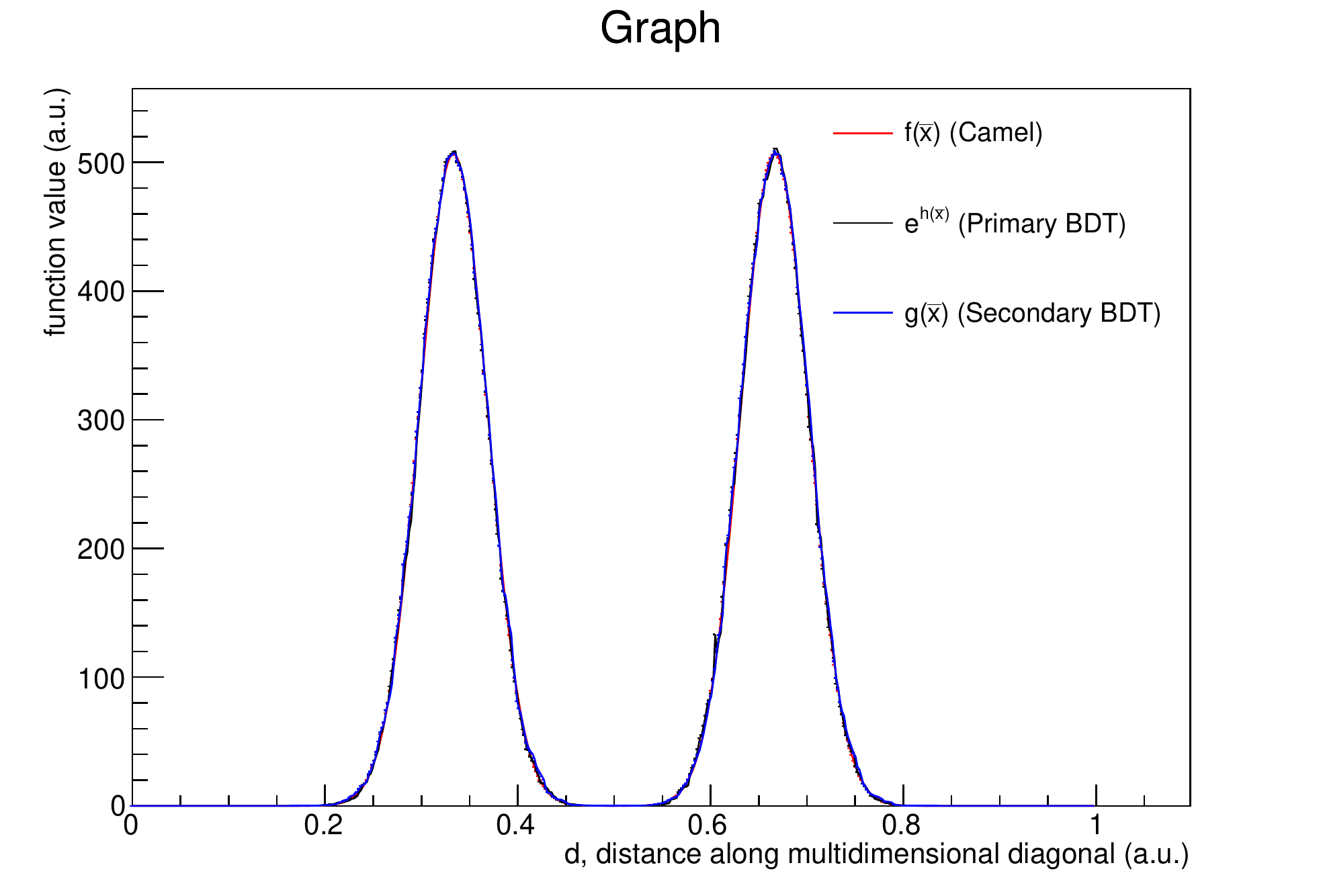}
 \includegraphics[width=0.5\textwidth]{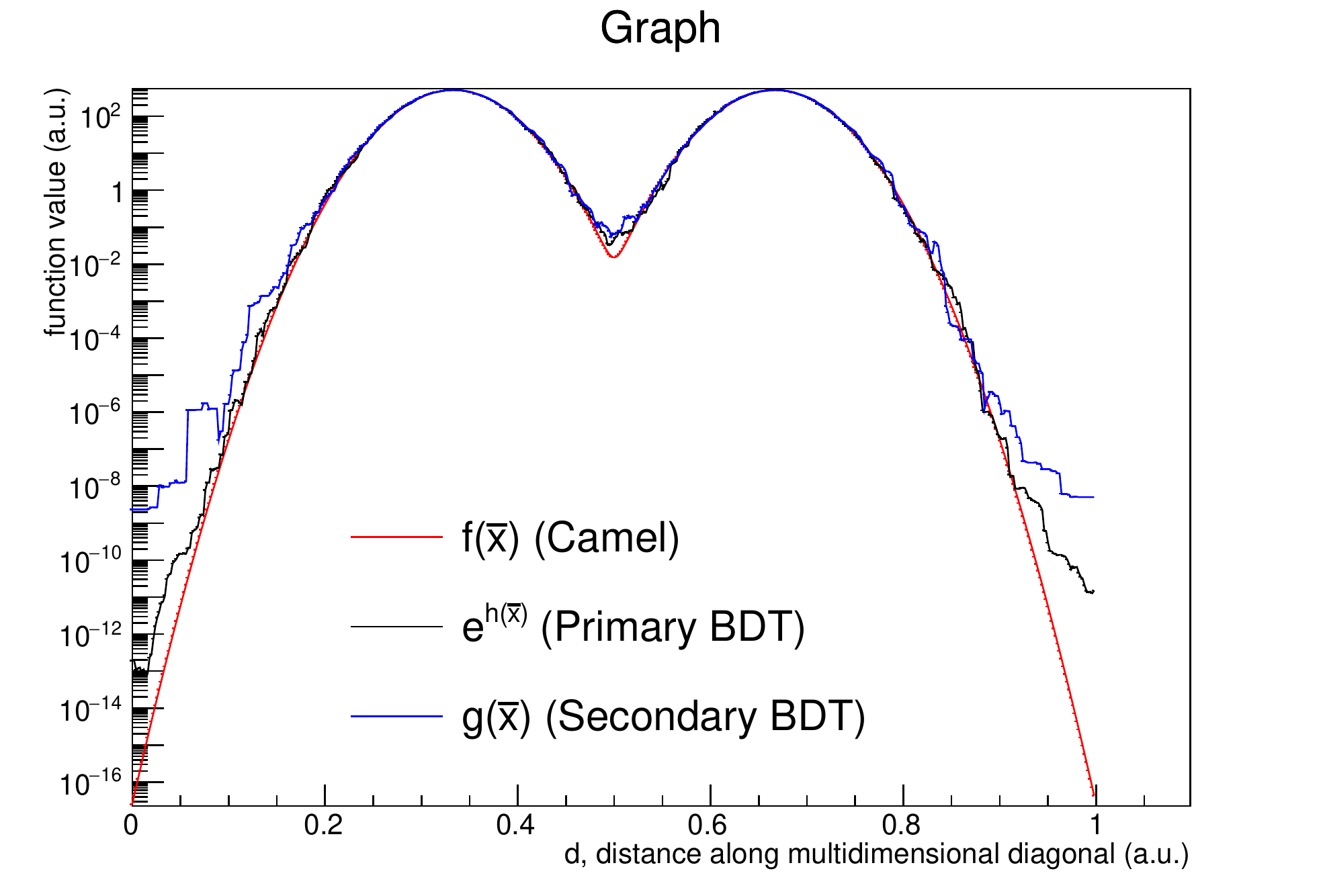}
 \caption{Comparison of the function value evaluated along the 4-dimensional diagonal on a linear (left) and logarithmic (right) scale between the target function $f(\bar x)$, the regression approximation $e^{h(\bar x)}$ and the generative BDT $g(\bar x)$.  While both the generative BDT and the regression approximation well-approximate the target function over many orders of magnitude, the generative BDT has a tendency to overestimate the function value in the very low-probability regions, with this effect somewhat mitigated by the regression approximation.}
 \label{fig:diag4d}
\end{figure}

Some diagnostic plots are shown for the 4-dimensional case, in which a total of $300,000$ function evaluations have been carried out during the training.  The function value evaluated along the multidimensional diagonal is shown in Figure \ref{fig:diag4d}, where the target function, regression approximation, and generative BDT are compared.  The tendency of the generative BDT to overestimate the target function value in regions of low probability is related to the non-negative constraint on the tree parameters, which makes it impossible to compensate for coarse-grained mapping of the function in these regions during early iterations of the BDT training.

\begin{figure}[htb!]
\includegraphics[width=0.5\textwidth]{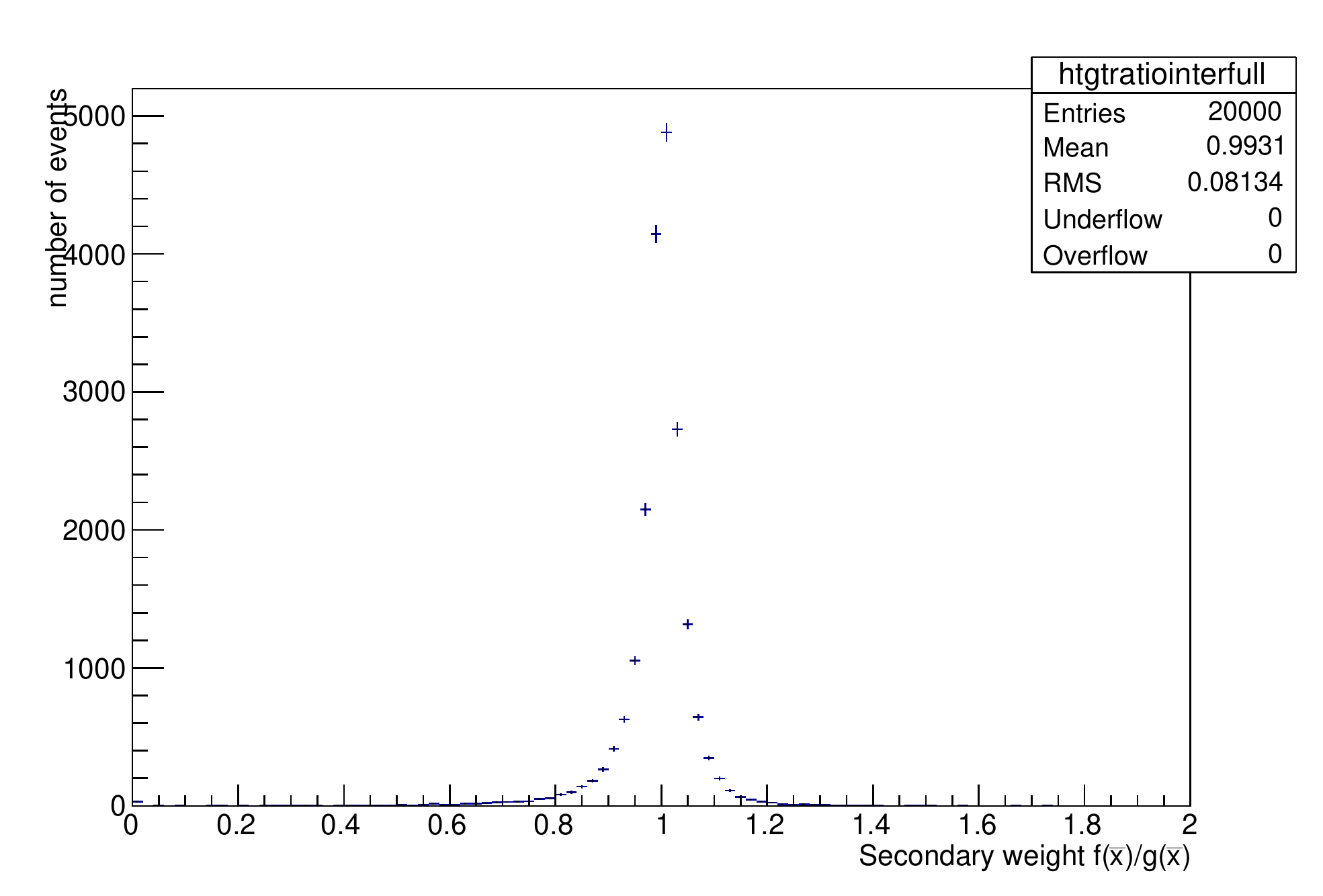}
\includegraphics[width=0.5\textwidth]{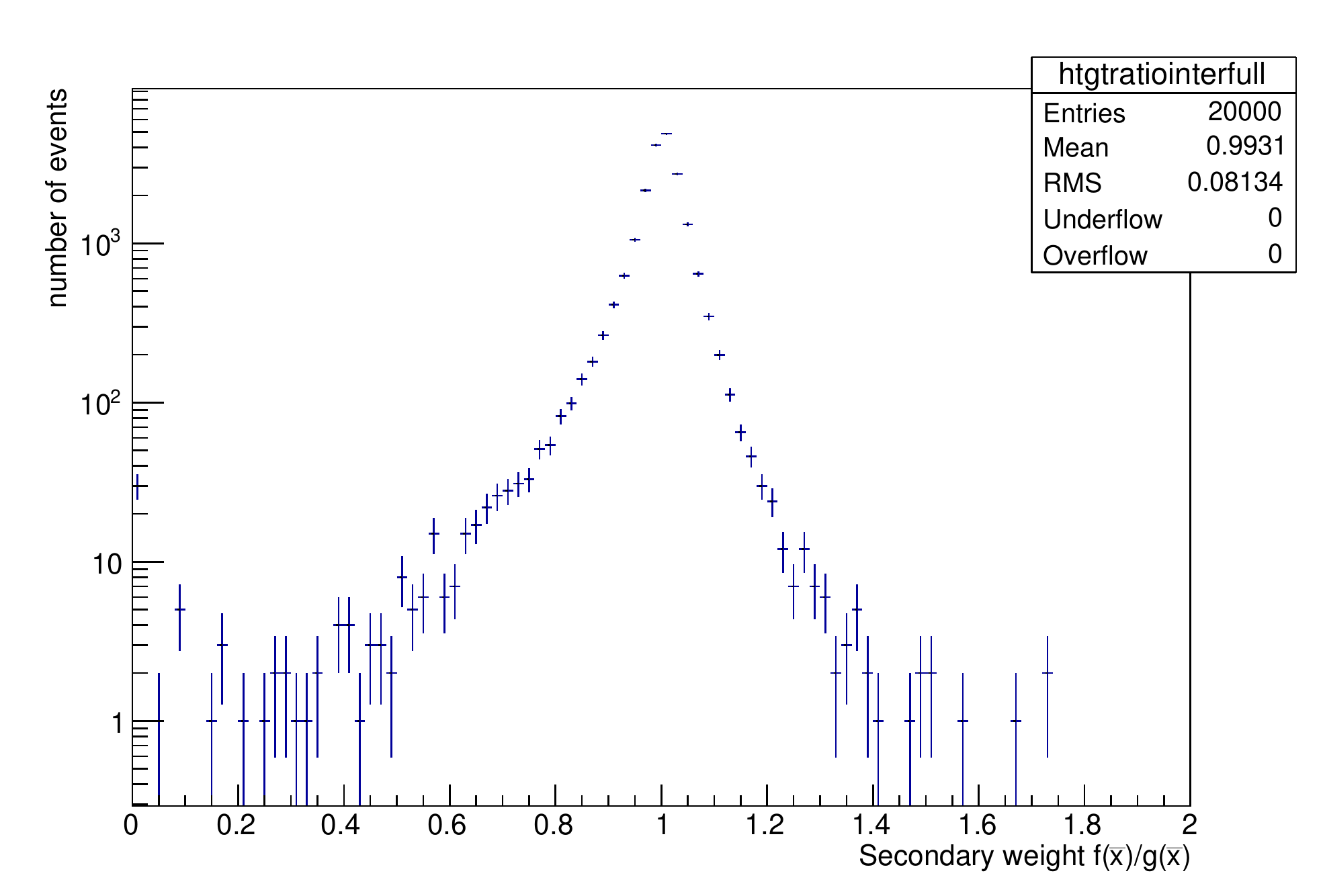}
\includegraphics[width=0.5\textwidth]{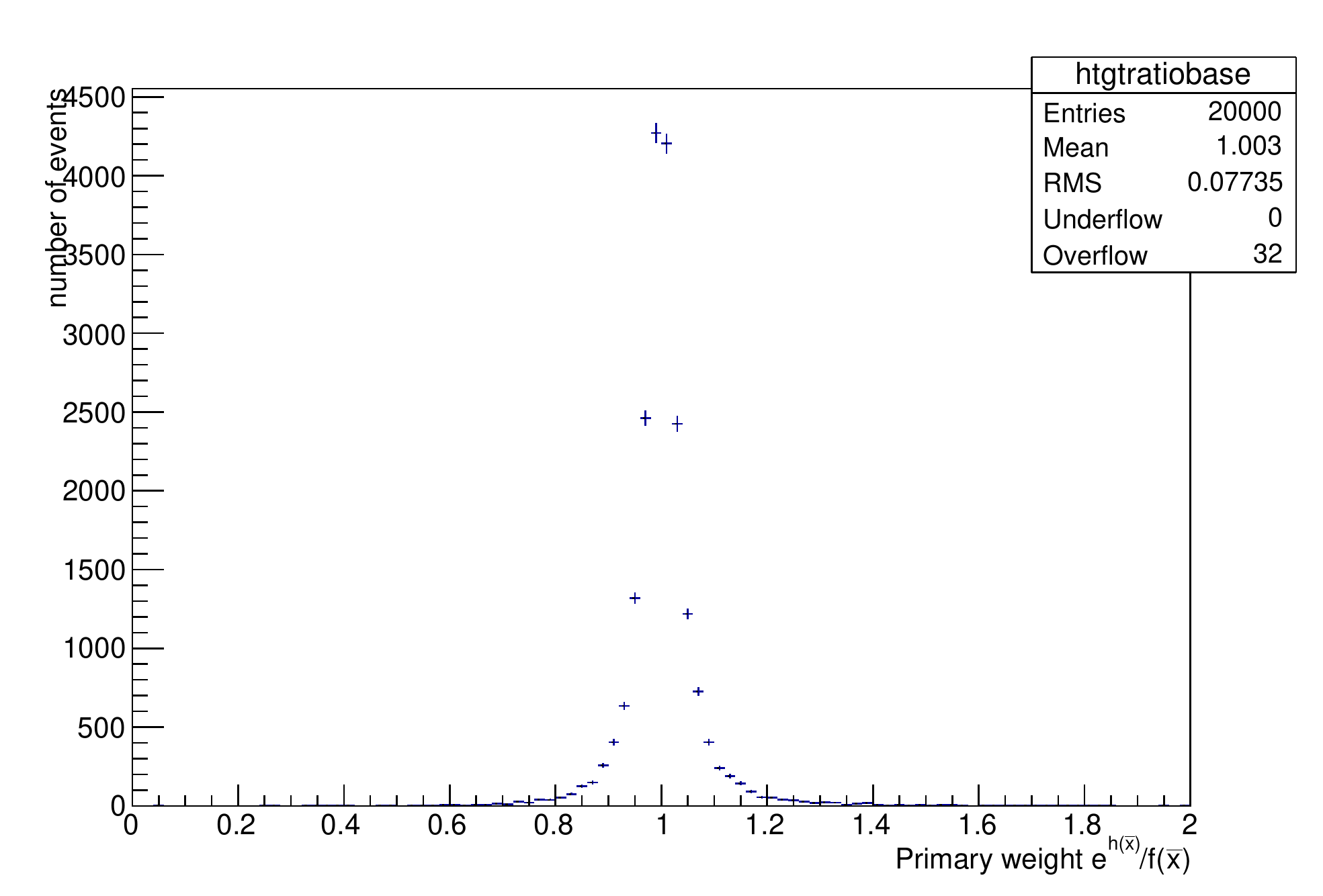}
\includegraphics[width=0.5\textwidth]{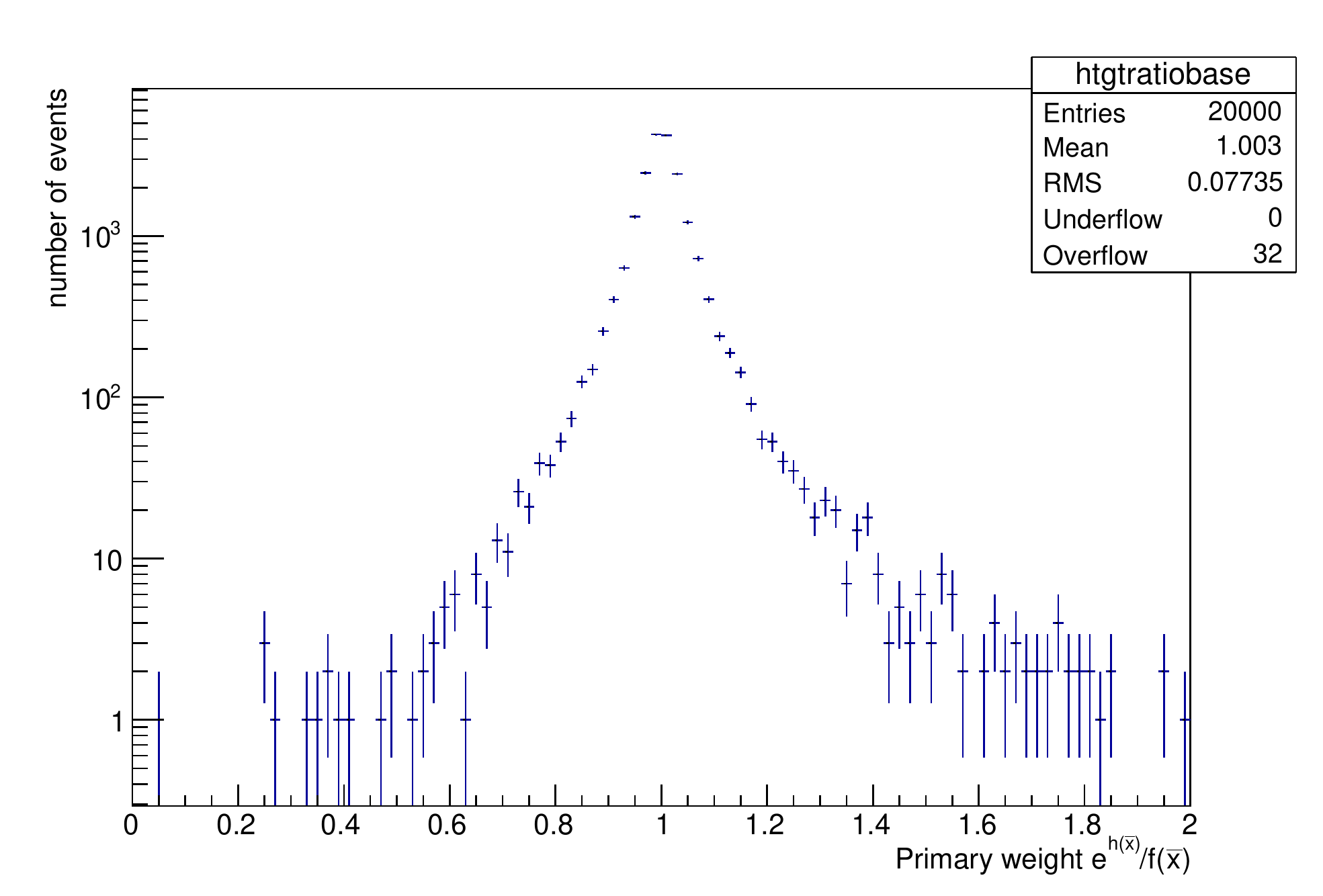}
\caption{The distribution of (non-normalized) integration weights for the 4-dimensional camel function are shown for the generative BDT (top) and the regression approximation (bottom), on linear(left) and logarithmic (right) scales.  Both distributions are reasonably symmetric, peaking at the true integral value, and with limited tails, with slightly smaller tails in the regression case.}
 \label{fig:intweights4d}
\end{figure}

The resulting non-normalized integration weight distributions $f(\bar x)/g(\bar x)$ and $f(\bar x)/e^{h(\bar x)}$ are shown in Figure \ref{fig:intweights4d}.

\begin{figure}[htb!]
 \includegraphics[width=0.5\textwidth]{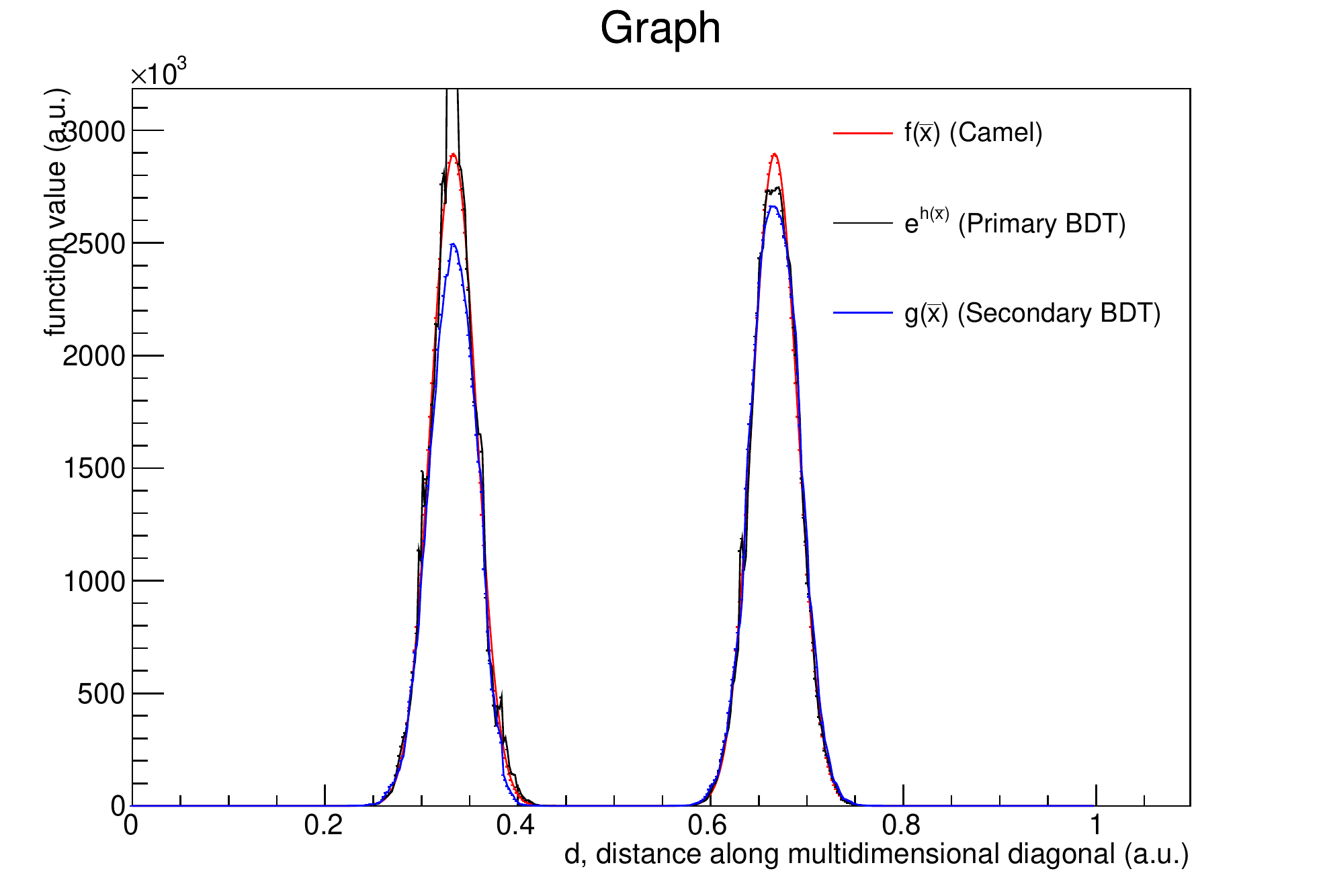}
 \includegraphics[width=0.5\textwidth]{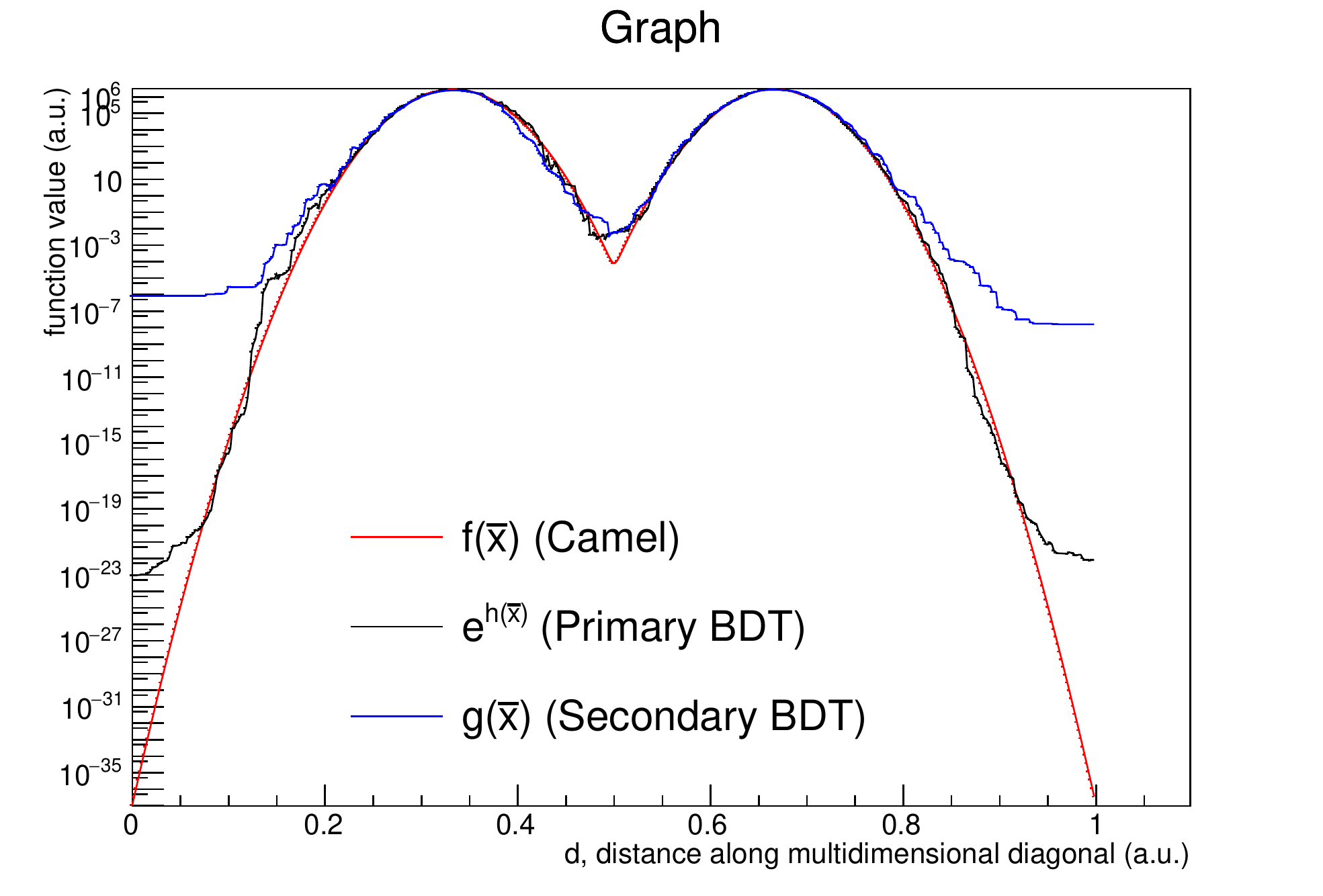}
 \caption{Comparison of the function value evaluated along the 9-dimensional diagonal on a linear (left) and logarithmic (right) scale between the target function $f(\bar x)$, the regression approximation $e^{h(\bar x)}$ and the generative BDT $g(\bar x)$.  The tendency of the generative BDT to overestimate the target function in low probability regions is increased with respect to the 4-dimensional case, while the regression is still well behaved in this regard.}
 \label{fig:diag9d}
\end{figure}

\begin{figure}[htb!]
\includegraphics[width=0.5\textwidth]{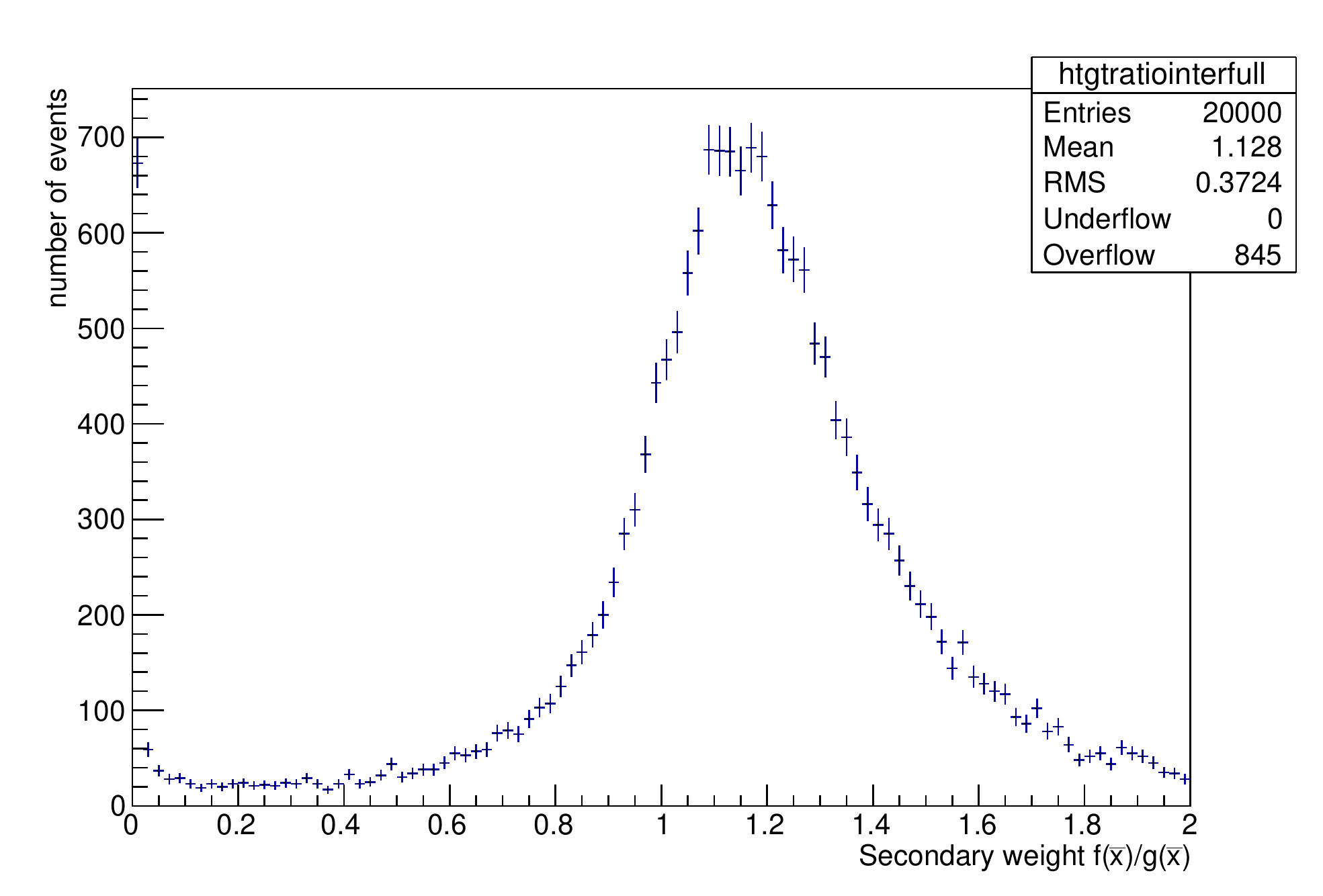}
\includegraphics[width=0.5\textwidth]{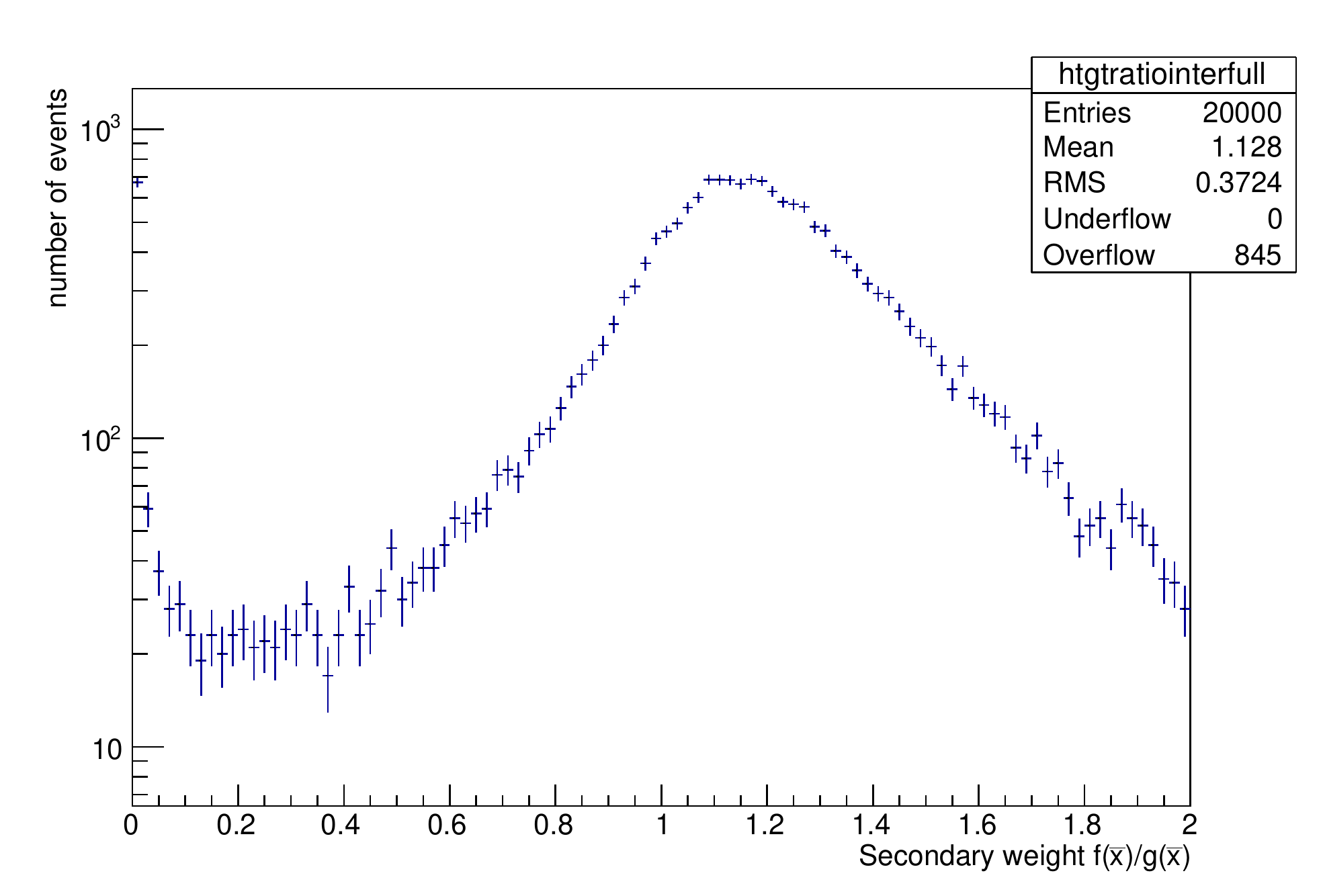}
\includegraphics[width=0.5\textwidth]{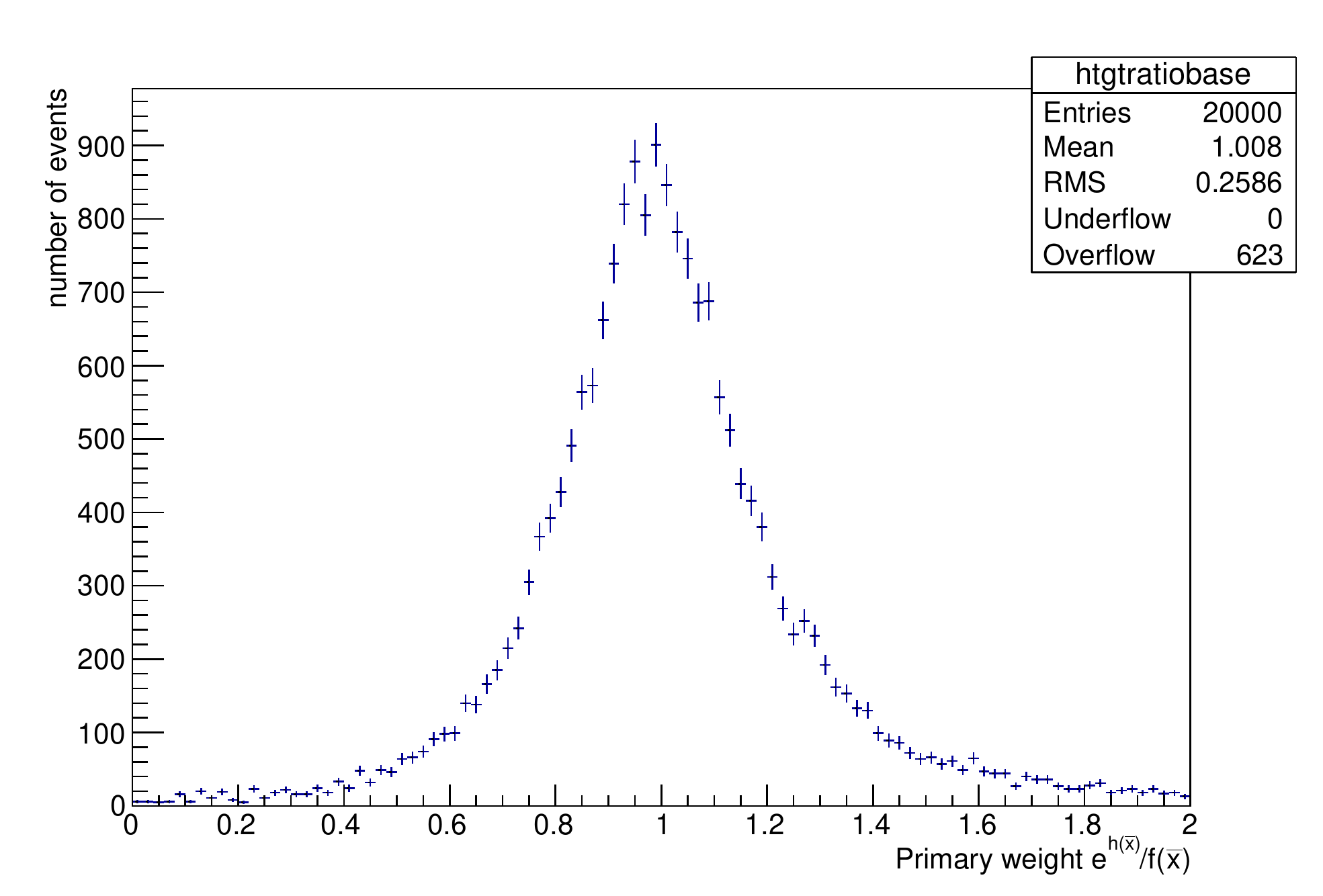}
\includegraphics[width=0.5\textwidth]{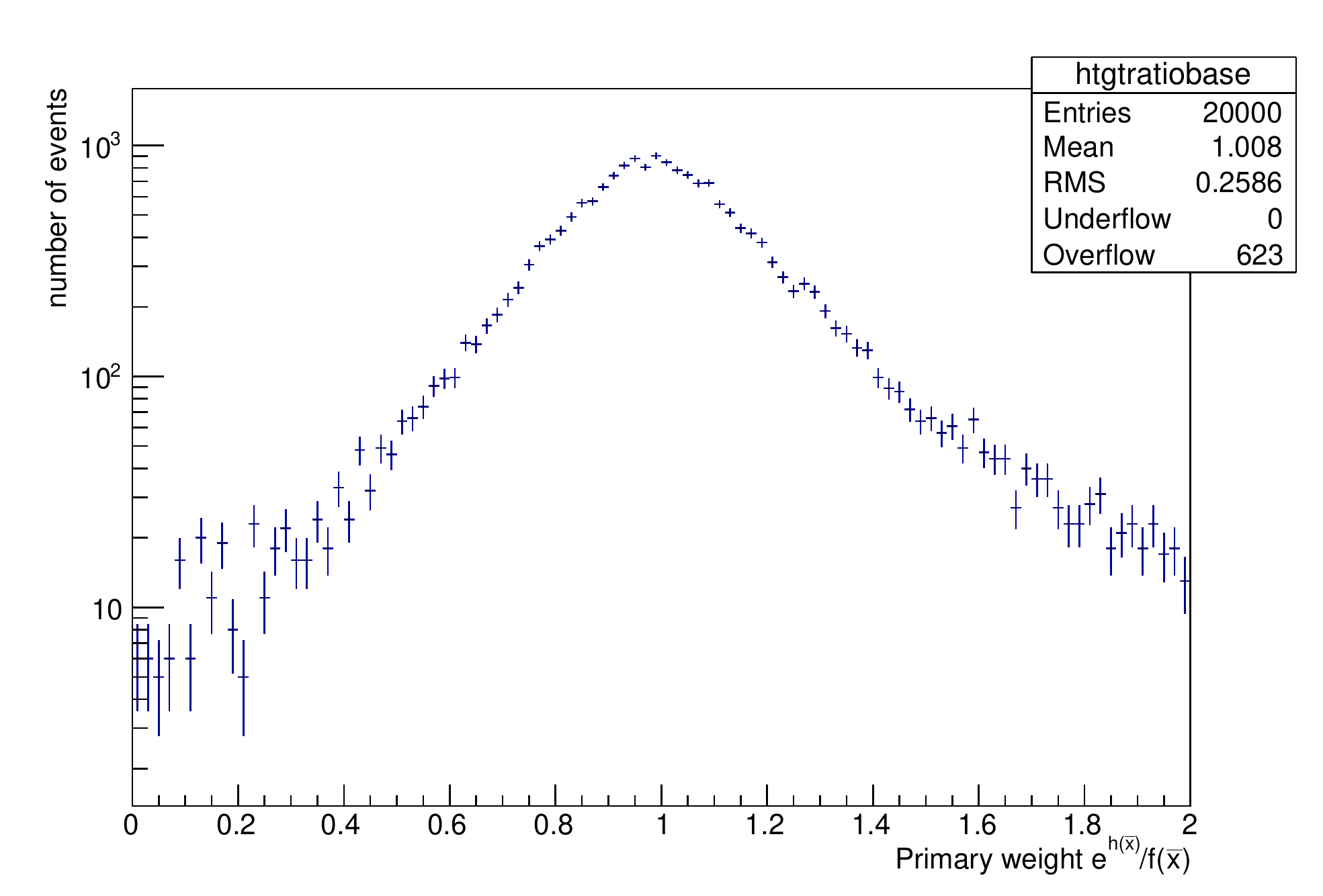}
\caption{The distribution of integration weights for the 9-dimensional camel function are shown for the generative BDT (top) and the regression approximation (bottom), on linear(left) and logarithmic (right) scales.  The overestimation of the target function by the generative BDT in low probability regions is visible here as a significant left tail, which is not present in the regression case.}
 \label{fig:intweights9d}
\end{figure}

The corresponding set of plots are shown in Figures \ref{fig:diag9d} and \ref{fig:intweights9d} for the 9-dimensional case, where a total of 3.2 million function evaluations have been used during the training.

\section{Generative Deep Neural Network Integration}
Recent work in data science and machine learning has had a significant focus on deep learning with artificial neural networks.  Several techniques have been developed for constructing generative models\cite{gan,vae,autoregressive}, often with a focus on image generation.  Most recent work on this topic has been carried out in the context of a fixed set of training data, or black box generator, for which a generative model is trained to reproduce similar examples, or samples following the same output distribution as the training set.  The problem for Monte Carlo integration with importance sampling is somewhat different, since in this case we cannot initially easily generate samples from the target distribution.  On the other hand, we are able to explicitly evaluate the value of the target probability density up to a normalizing constant.\\

We start from the paradigm used for generative adversarial networks\cite{gan}, in which data is sampled from an easily computable and sampled prior distribution $p(\bar z)$, and passed through a generative model $G$ which maps a sample from the prior $\bar z$ onto an output $\bar x$, where in general $\bar z$ and $\bar x$ need not have the same dimensionality, and in typical image generation cases the dimensionality of $\bar z$ is much smaller than that of $\bar x$.  The transformation $G$ is typically given by a deep neural network, often with some convolutional layers in the image generation case.

For Monte Carlo integration we instead enforce the same dimensionality for $\bar z$ and $\bar x$, and exploit the fact that in this case the prior probability density is related to the probability density of the generated samples $p_g(\bar x)$ according to
\begin{equation}
 p(\bar z) = p_g(\bar x)\left|\left|\frac{\partial \bar G(\bar z)}{\partial\bar z}\right|\right|
 \label{eqn:pg}
\end{equation}
where $\frac{\partial \bar G(\bar z)}{\partial\bar z}$ is a $d$ by $d$ matrix which is the Jacobian of the transformation $G$, and $\left|\left|\frac{\partial \bar G(\bar z)}{\partial\bar z}\right|\right|$ is the absolute value of the determinant of the Jacobian.  The optimal Monte Carlo integration precision would be achieved for $p_g(\bar x) \approx p_f(\bar x)$, where $p_f(\bar x) = f(\bar x)/I_f$ is the normalized probability density corresponding to the target function $f(\bar x)$, and therefore we would like to find a loss function which can be minimized in order to achieve this.\\

The KL-divergence\cite{kl} from $p_f(\bar x)$ to $g(\bar x)$ is defined as
\begin{equation}
 D_{KL} = \int g(\bar x)\ln\frac{p_g(\bar x)}{p_f(\bar x)}d\bar x
\end{equation}
and is a measure of the difference between the two probability distributions, taking on a minimum value of $0$ in the case where the two distributions are identical.  This KL divergence can be approximated numerically from a finite data set of size $N$ sampled from the prior $p(\bar z)$ according to
  \begin{equation}
  D_{KL} = \frac{1}{N}\sum \left[\ln p(\bar z) - \ln\left|\left|\frac{\partial \bar G(\bar z)}{\partial\bar z}\right|\right| - \ln f(G(\bar z)) \right] + I_f
  \label{eqn:directloss}
  \end{equation}
where $I_f$ is a constant and can be neglected, which is fortunate, since it is not known a priori.  If $G$ is a deep neural network with $d$ inputs and $d$ outputs and suitably continuous and well behaved activation functions\footnote{Contrary to typical loss functions, the first derivatives of the activation functions appear directly in the loss function, and therefore any activation functions which are used must have continuous first derivatives, excluding the use of rectifying linear units for example.}, the above can be used directly as a differentiable\footnote{The matrix determinant appearing in this expression is normally computed from a non-differentiable matrix decomposition, but its derivative can be evaluated from Jacobi's formula according to $\frac{\partial}{\partial t}\ln\left|\left|A\right|\right| = tr\left(A^{-1}\frac{\partial}{\partial t}A\right)$} loss function in stochastic gradient descent (SGD) \textit{provided that the target function $f(\bar x)$ is easily computed and differentiable}.\\

In the typical case where $f(\bar x)$ and/or its derivatives are difficult or computationally expensive to evaluate, we can train an additional regression to approximate it.  Since this regression output will appear in the loss function used to optimize the generative model $G$ with SGD, it must be continuous, ruling out a BDT regression which is only piecewise continuous.  A regression using deep neural networks satisfies the continuity requirement on the other hand, and is also convenient from an implementation standpoint since it can be implemented in the same software framework as the generative model.  Similar to the BDT case we introduce a regression function $h(\bar x)$ such that $e^{h(\bar x)} \approx f(\bar x)$.  Imposing the desired log-normal constraint on $e^{h(\bar x)}$, this can again be trained with the loss function in (\ref{eqn:lognorm}).  In this case the loss function for the generative model can be written replacing $f(\bar x)$ with $e^{h(\bar x)}$.
  \begin{equation}
  D_{KL} = \frac{1}{N}\sum\left[\ln p(\bar z) - \ln\left|\left|\frac{\partial \bar G(\bar z)}{\partial\bar z}\right|\right| - h(G(\bar z)) \right]
  \end{equation}

In order to train the regression and generative models, an iterative procedure\footnote{It should be noted that the iterative procedure defined here differs significantly from the GAN case.  Since this is not a MINIMAX problem, there are no strong equilibrium requirements between $G$ and $h$, and both can be trained to convergence at each iteration of the procedure without the need to switch back and forth between them many times as in the GAN training.  This also means that most of the stability issues with GAN training are avoided.} can be defined similar to the BDT case.

\begin{enumerate}
 \item Create an initial training dataset by sampling from an easily sampled prior distribution (by default a uniform distribution over the specified integration integration range) and evaluating the target function value $f(\bar x)$ for each sampled point.
 \item Iteratively sample and train DNN's:
 \begin{enumerate}
  \item Train regression model $h(\bar x)$
  \item Train generative model $G(\bar z)$ (keeping model parameters of $h(\bar x)$ fixed)
  \item Check for convergence and end training if appropriate
  \item Add to or replace the training dataset by sampling additional events from the prior $p(\bar z)$, transforming them with the generative model $G(\bar z)=\bar x$ and evaluating target function $f(\bar x)$ for each sampled point
 \end{enumerate}
\end{enumerate}

In case the target function is zero or undefined in some regions within the integration range, the loss functions in (\ref{eqn:lognorm}) or (\ref{eqn:directloss}) would be undefined.  For the case of the regression loss function, this could be modified simply with a numerical cutoff.
\begin{equation}
 L_h = \sum\left(\ln \textnormal{max}(f(\bar x),\epsilon) - h(G(\bar z))\right)^2
\end{equation}

The same approach could be used for the generative loss function in case no intermediate regression is used.
  \begin{equation}
  D_{KL} = \frac{1}{N}\sum\left[\ln p(\bar z) - \ln\left|\left|\frac{\partial \bar G(\bar z)}{\partial\bar z}\right|\right| - \ln \textnormal{max}(f(G(\bar z)),\epsilon) \right] + I_f
  \end{equation}

Alternatively, a modified loss function can be defined related to the standard binary cross-entropy loss function for a classifier, closely related to the loss function in \cite{gan}.
  \begin{equation}
  D = \frac{1}{N}\sum\left[\ln p(\bar z) - \ln\left|\left|\frac{\partial \bar G(\bar z)}{\partial\bar z}\right|\right| - \ln\left(\frac{p(\bar z)}{\left|\left|\frac{\partial \bar G(\bar z)}{\partial\bar z}\right|\right|} + \frac{f(G(\bar z))}{I_f}\right) \right]
  \end{equation}

Similarly this alternative loss function can also be used in conjunction with the regression approximation $e^{h(\bar x)}$.
  \begin{equation}
  D = \frac{1}{N}\sum\left[\ln p(\bar z) - \ln\left|\left|\frac{\partial \bar G(\bar z)}{\partial\bar z}\right|\right| - \ln\left(\frac{p(\bar z)}{\left|\left|\frac{\partial \bar G(\bar z)}{\partial\bar z}\right|\right|} + \frac{e^{h(G(\bar z))}}{\int e^{h(\bar x)}d\bar x}\right) \right]
  \end{equation}

This loss function has the disadvantage that the integral $I_f$ or $\int e^{h(\bar x)}d\bar x$ must be known.  In practice, a reasonable approximation of this integral is sufficient, which can be computed with no additional function evaluations from the training samples generated at each iteration of the training.


Once the training is complete, numerical integration of the target function can be performed by sampling an arbitrary number of phase-space points from the prior $p(\bar z)$ and transforming them with the generative model $G(\bar z)$ until the desired precision is reached, where as in the BDT case, the value of the integral is given by the mean of the integration weights $<f(\bar x)/p_g(\bar x)>$ and the uncertainty is given by (\ref{eqn:uncert}).  The generating PDF $p_g(\bar x)$ can be explicitly evaluated for each generated phase-space point according to (\ref{eqn:pg}).

As in the BDT case, but for different underlying reasons, the regression DNN is in general easier to train and can achieve a better precision compared to the generative model, and so the staged approach can also be carried out here, where the generative model is used to evaluate the integral of the regression approximation $e^{h(\bar x)}$ and to sample from it using accept-reject sampling with reasonable efficiency, with those samples then used to evaluate the integral of the target function.

\subsection{Implementation Details}
Training and generation for the DNN-based models are implemented in python using a combination of Numpy\cite{numpy}, Tensorflow\cite{tensorflow}, and Keras\cite{keras}.  Tensorflow is used both as the backend for Keras, and also to directly implement the logarithm of the absolute value of the determinant which appears in the loss function for the generative model.  In this context, an additional Tensorflow Op is implemented in order to compute the logarithm of a matrix determinant directly, rather than chaining up existing logarithm, absolute value, and matrix determinant operations, in order to avoid numerical issues which can arise in case of extreme values.  While Tensorflow supports running on both CPU's (with multi-threading) and GPU's, there are not yet GPU implementations for the log matrix determinant, nor for the matrix inverse operation needed to calculate its gradient.  The built-in matrix determinant operation in Tensorflow is also lacking a GPU implementation, and also relies on the matrix inverse for its gradient.

The generative and regression models are currently implemented as fully connected neural networks with 5 hidden layers each.  The generative model is using a modified hyperbolic tangent activation function for the hidden layers $0.7\ \textnormal{tanh}(\bar x) + 0.3\ \bar x$, which ensures both continuous first derivatives, and that the output range is $[-\infty,\infty]$.  The output layer uses sigmoid activation functions in order to restrict the generated phase space points to a unit hyper-cube.  This could of course be trivially shifted and/or scaled to accommodate an alternate integration range.  Combined with the output range of the hidden layers, this formally ensures that the generative model has full support over the integration range.  The regression model is rather using exponential linear units (ELU) for the activation function on the hidden layers, and a linear activation on the output layer.  For the examples shown here, the generative model is using 64 nodes per hidden layer, and the regression model is using 32.

In order to ensure stable convergence in particular for the generative model, it is required to use relatively large batch sizes, with 5120 used here, as well as a sufficiently small learning rate and gradient clipping.  In order to efficiently train the model, the number of training epochs is selected dynamically by using the early stopping functionality in Keras.  In order to ensure a sufficiently small learning rate near convergence while efficiently training the model at early epochs, the ``ReduceLROnPlateau'' functionality in Keras is used to dynamically reduce the learning rate at each epoch where the loss function stops decreasing over the past several iterations.


\subsection{Illustrative Example}
In order to further examine key features of this approach, it is useful to look at simple one-dimensional examples with known analytic solutions.  The training procedure has been carried out for a target function corresponding to a one-dimensional Cauchy distribution
\begin{equation}
 f(\bar x) = \frac{1}{\pi}\frac{\Gamma}{\left(x-\frac{1}{2}\right)^2+\Gamma^2}
\end{equation}
with $\Gamma=1/10$ and $x\in[0,1]$.  One important feature is that in one dimension, an exact generative model $G_a(z)=x$ can always be written in terms of the cumulative distribution function (CDF) of the prior $p(z)$ and the inverse CDF of the target density $p_f(x)$, following the usual inverse transform sampling procedure.
\begin{equation}
 G_a(z) = \textnormal{CDF}^{-1}_{p_f}\left[\textnormal{CDF}_p(z)\right]
\end{equation}

In this particular case there is a closed form analytic solution.
\begin{equation}
  G_a(z) = \frac{1}{2} + \Gamma\tan\left[\arctan\left(\frac{1}{2\Gamma}\right)\textnormal{erf}\left(\frac{z}{\sqrt{2}}\right)\right]
\end{equation}

\begin{figure}[htb!]
\begin{center}
\includegraphics[width=0.5\textwidth]{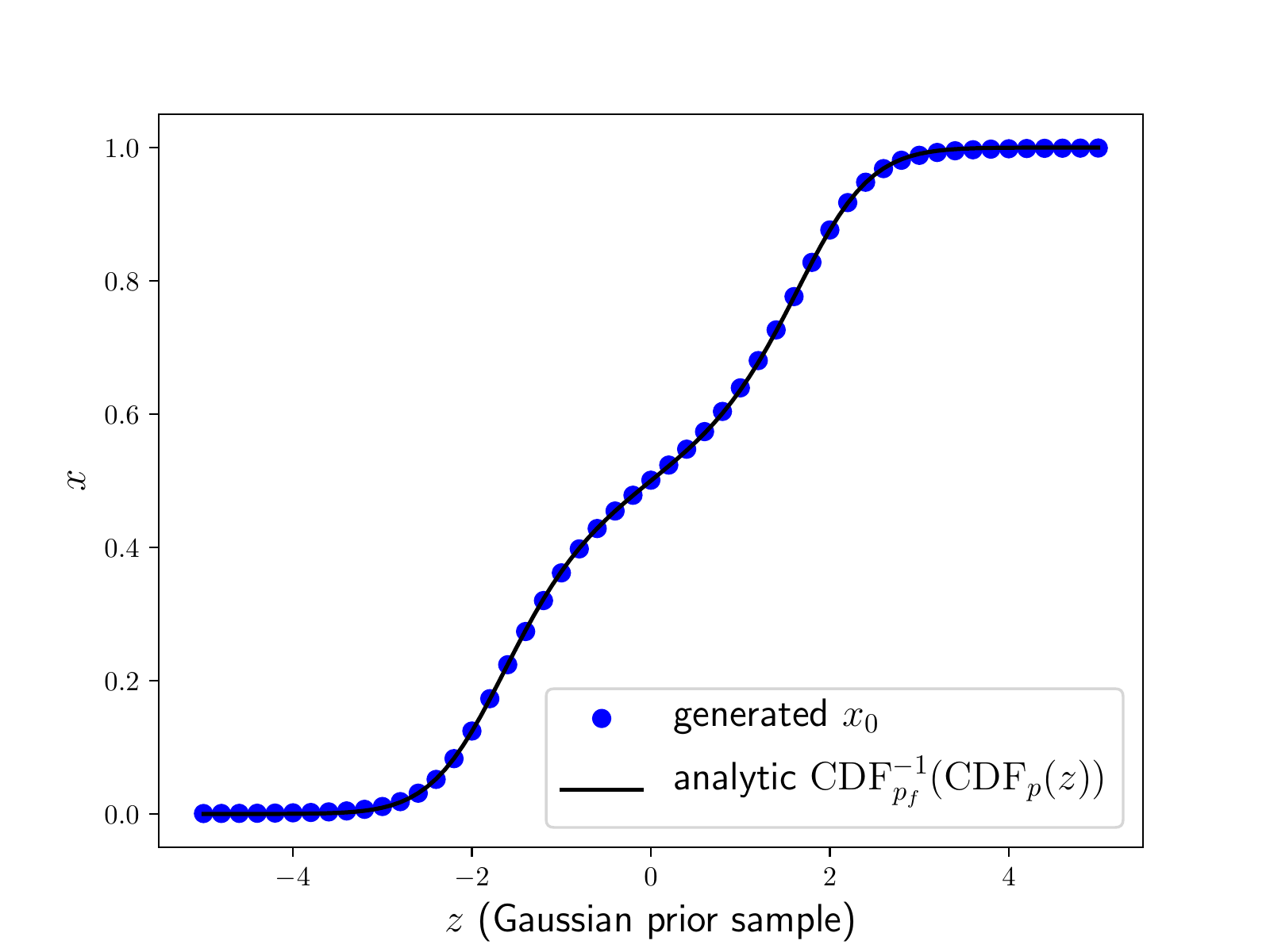}
\end{center}
\caption{The output of the generative model $x$ as a function of the input $z$.  The trained generative model is compared to the known analytic solution for this example, with excellent agreement.}
 \label{fig:cauchygen}
\end{figure}

The comparison between the analytic solution and the trained generative model is shown in Figure \ref{fig:cauchygen}.  The good agreement between the trained generative model and the analytic solution indicates that generative model is able to learn the inverse CDF numerically during the training procedure\footnote{In fact there is an ambiguity given the by the absolute value of the determinant which appears in the loss function.  In one dimension this would imply that $G(z) = \textnormal{CDF}^{-1}_{p_f}\left[1-\textnormal{CDF}_p(z)\right]$ is an equally valid solution.  Since the prior density is symmetric around zero in $z$, this is equivalent of transforming $z\rightarrow -z$ and corresponds to a change in sign of the determinant, which has no effect on the KL divergence due to the absolute value.  In the multi-dimensional case, this ambiguity can in fact be manifested by any rotation or reflection around the origin in $\bar z$, since this would in general leave the magnitude of the determinant unchanged.}.  This suggests that the generalization of this procedure to multiple dimensions can be considered as a variation of inverse transform sampling, where the required multidimensional transformation is inferred from deep learning methods.

\begin{figure}[htb!]
 \includegraphics[width=0.5\textwidth]{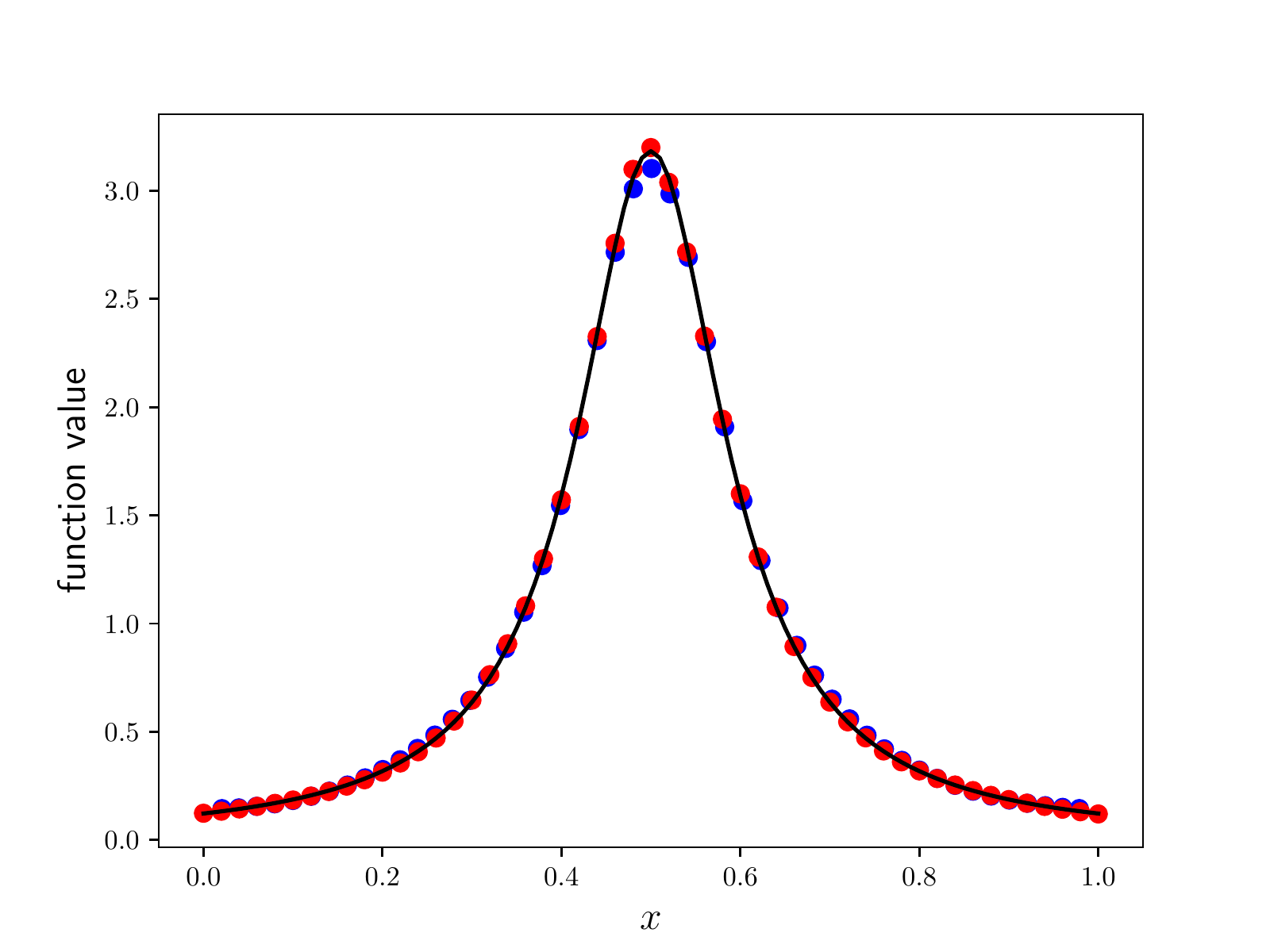}
 \includegraphics[width=0.5\textwidth]{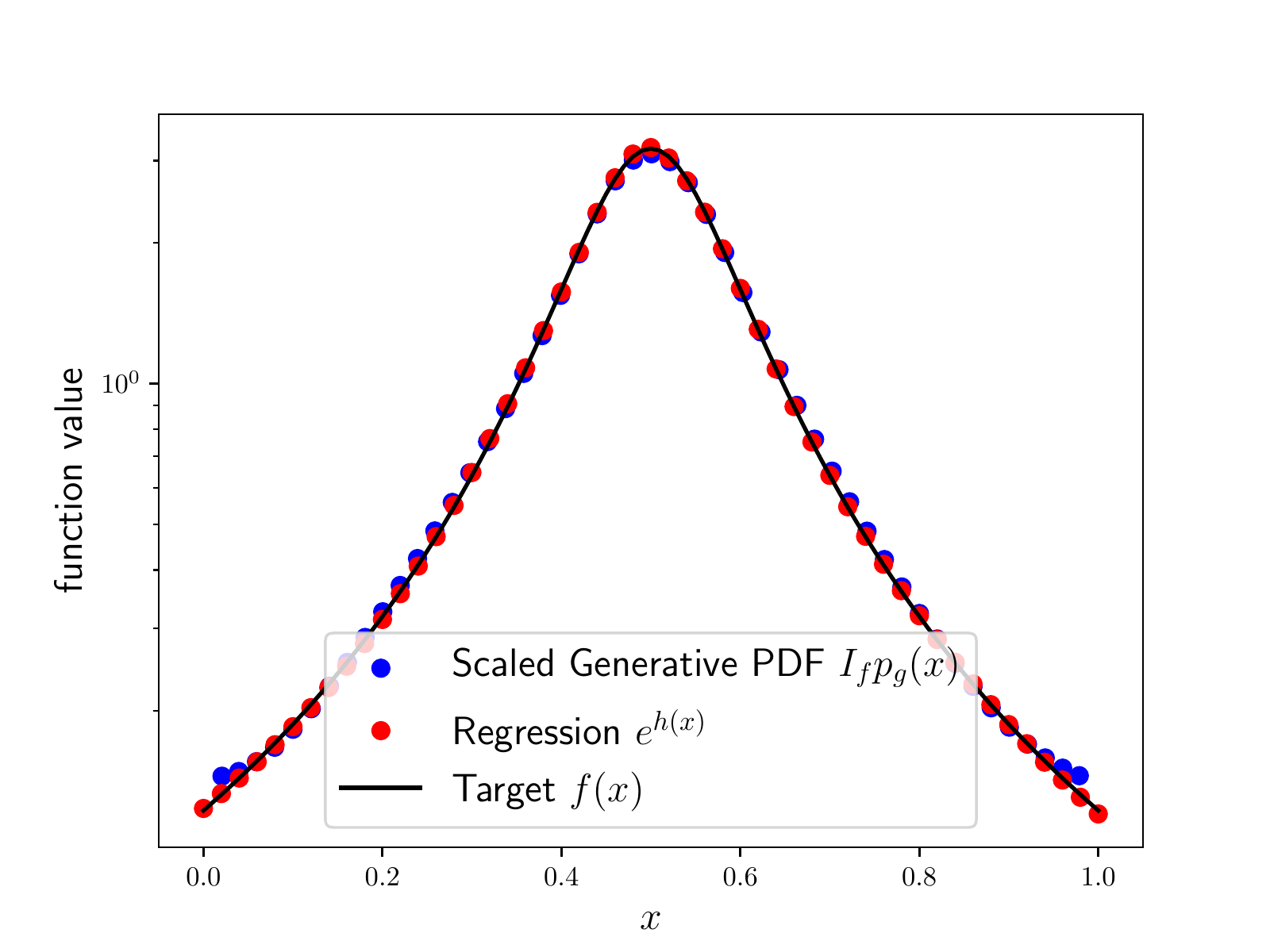}
 \caption{Comparison of the target function value with the corresponding approximations from the regression and generative models on a linear (left) and logarithmic (right) scale.}
 \label{fig:cauchyreg}
\end{figure}

The behaviour of both the regression and generative models can also be compared directly to the target function value, with the regression approximation given by $e^{h(\bar x)}$ and the function value implied by the generative model given by $I_f p_g(\bar x)$.  This comparison is shown in Figure \ref{fig:cauchyreg}.  An important feature is that the probability density $p_g(\bar x)$ associated with the generative model cannot in general be evaluated for arbitrary phase space points $\bar x$, but only those for which the corresponding phase space point $\bar z$ from the input space is known.  This limitation is not crucial for Monte Carlo integration or unweighting, since in these cases all of the phase space points are sampled from the generative model itself, and therefore $\bar z$ is always known.  Evaluating $p_g(\bar x)$ for arbitrary phase space points requires computing the inverse of the generative model.  This should exist as long as there is a 1:1 mapping between $\bar z$ and $\bar x$ and $p_g(\bar x)$ has full support over $\bar x$, but the invertibility has not yet been studied in detail.

\begin{figure}[htb!]
 \includegraphics[width=0.5\textwidth]{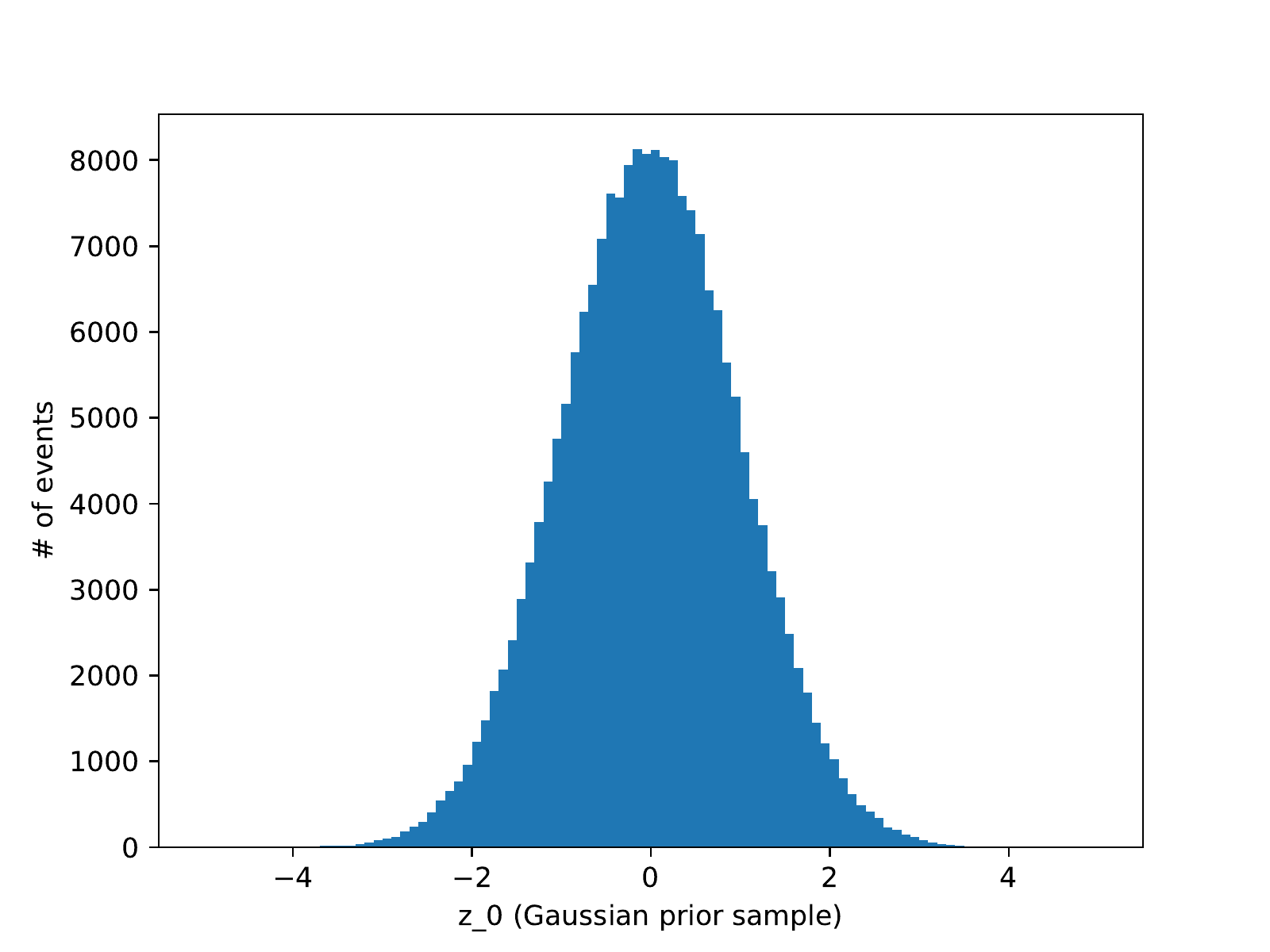}
 \includegraphics[width=0.5\textwidth]{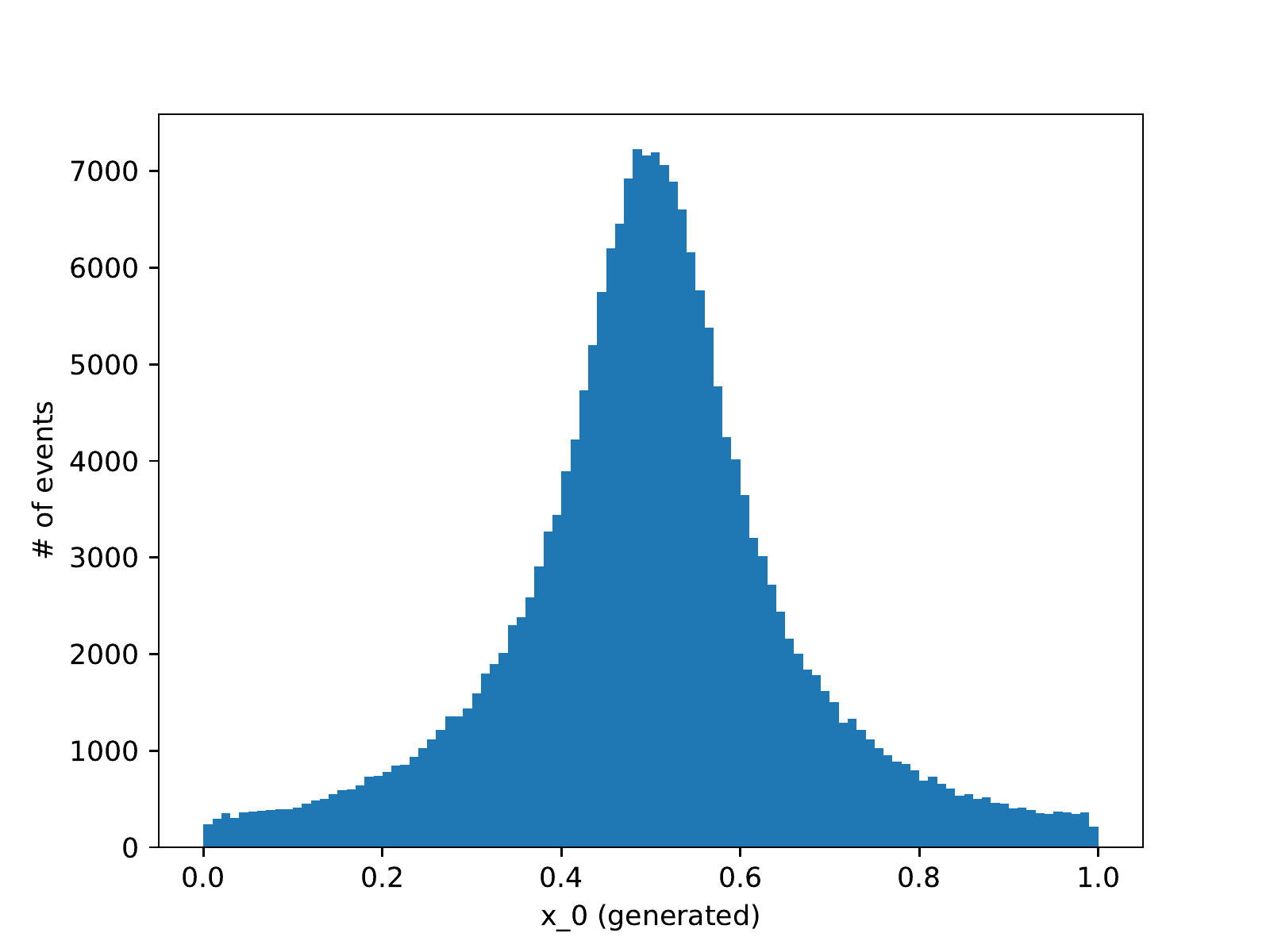}
 \caption{The distribution sampled from the unit gaussian prior (left) and after transformation by the generative network (right).}
 \label{fig:cauchygend}
\end{figure}

The distribution sampled from the prior is shown before and after being transformed by the generative model in Figure \ref{fig:cauchygend}.


\subsection{Results}

\begin{figure}[htb!]
 \includegraphics[width=0.5\textwidth]{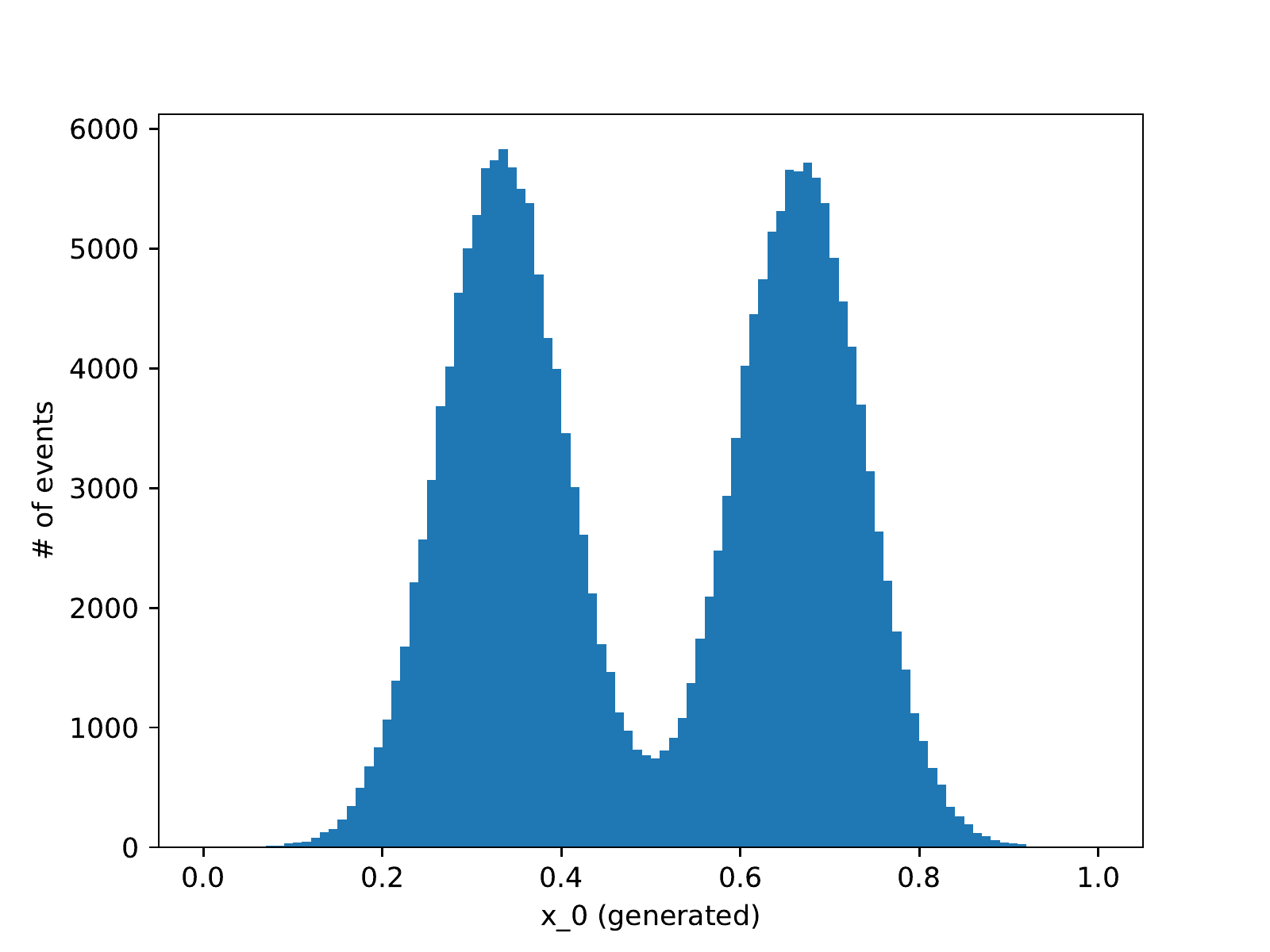}
 \includegraphics[width=0.5\textwidth]{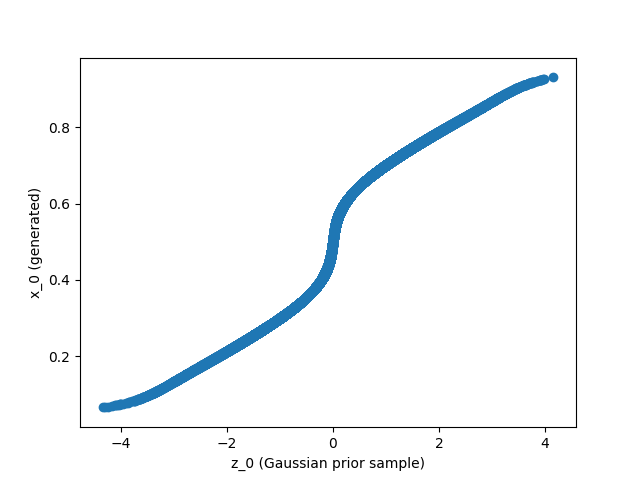}
 \caption{Generated distribution for the generative model trained against the 1-dimensional camel distribution (left) and the generated output value as a function of the input (right).}
 \label{fig:camel1d}
\end{figure}

As in the BDT case, performance test are carried out using the multi-dimensional camel function.  Firstly as a reference point, some basic plots are shown for the 1-dimensional camel function in Figure \ref{fig:camel1d}, showing the generated distribution, and the generated value of $x$ as a function of the prior $z$.  No closed form analytic solution is available to compare to in this case.

\begin{figure}[htb!]
\includegraphics[width=0.5\textwidth]{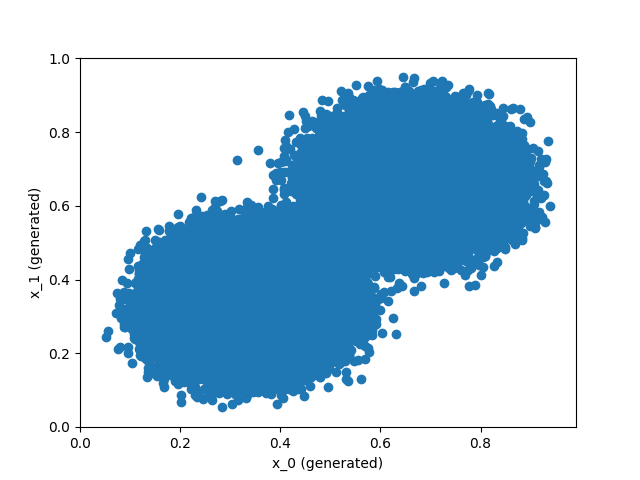}
\includegraphics[width=0.5\textwidth]{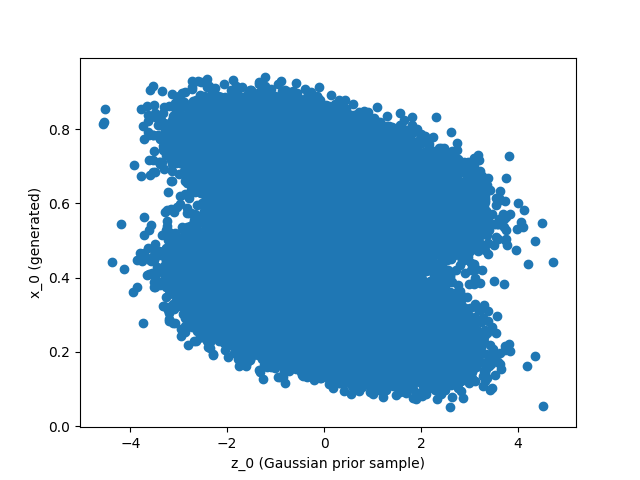}
\caption{Two-dimensional projection of the samples generated by the generative model trained against the 4-dimensional camel function (left) and correlation between output value $x_0$ and input value $z_0$ (right).  Since the generative model is not factorized, a single output $x_0$ depends in general on all four of the values in the input vector $\bar z$ and not just $z_0$, such that the relationship between $x_0$ and $z_0$ is not 1:1, even if the relationship between $\bar x$ and $\bar z$ remains so.}
 \label{fig:4dcamelgen}
\end{figure}

Diagnostic plots are shown for the 4-dimensional case, in which a total of $294,912$ function evaluations have been carried out during the training.  Two-dimensional projections of the samples from the generative model and correlation between generated samples and input values are shown in Figure \ref{fig:4dcamelgen}.

\begin{figure}[htb!]
 \includegraphics[width=0.32\textwidth]{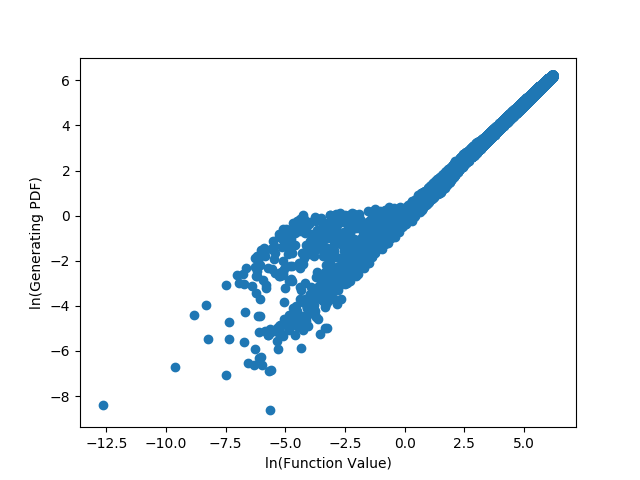}
 \includegraphics[width=0.32\textwidth]{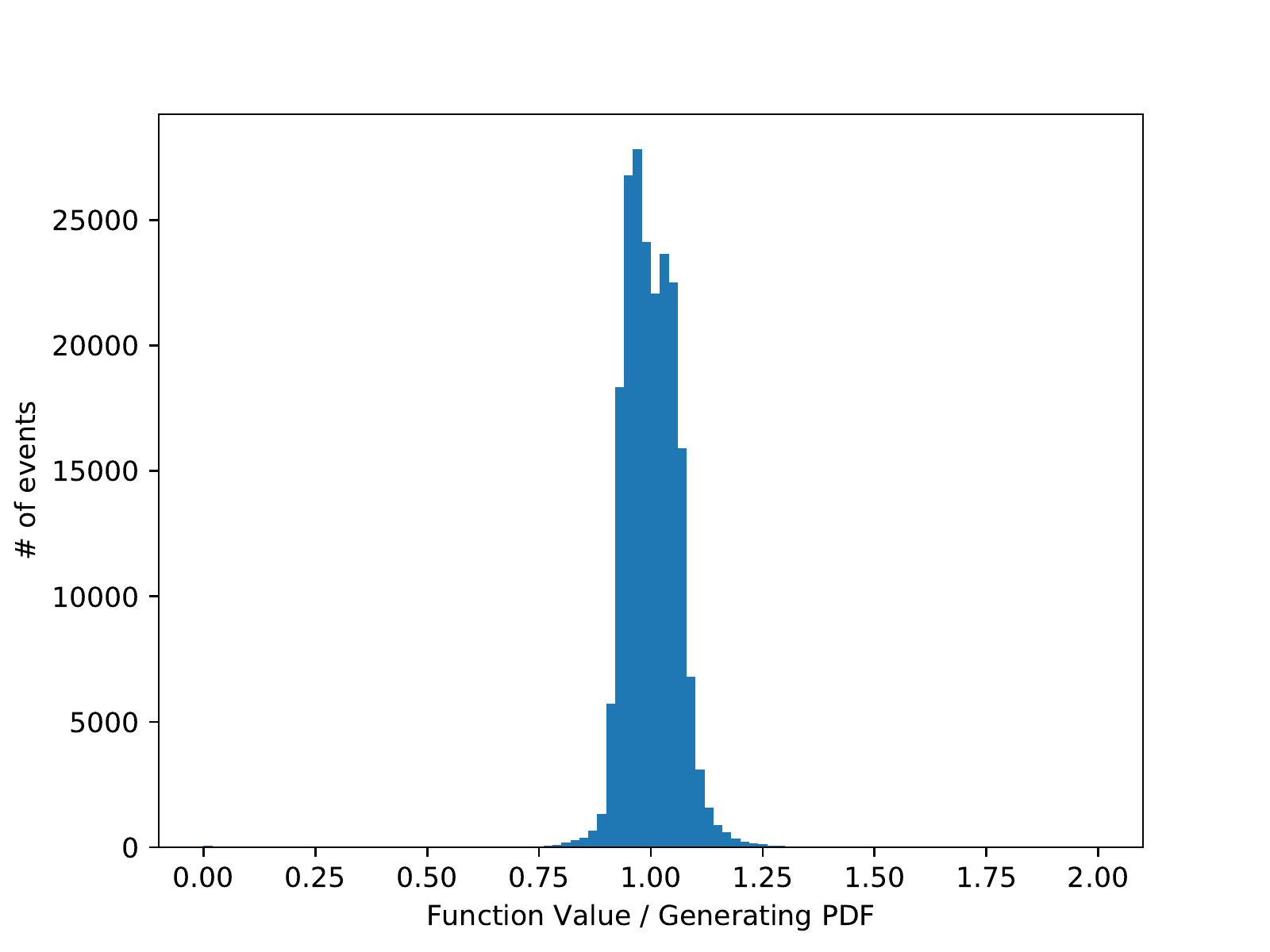}
 \includegraphics[width=0.32\textwidth]{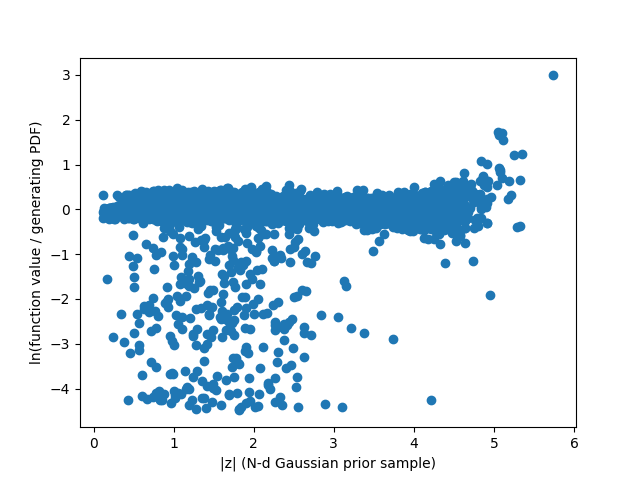}
 \caption{The correlation is shown between the PDF associated with the generative model $p_g(\bar x)$ and the target function value $f(\bar x)$ for the 4-dimensional camel function (left).  The generating PDF tracks the target function down to relatively low values, but tends to overestimate the target in regions of very low probability.  The corresponding integration weight distribution is shown (center), as well as the correlation of the integration weight with the magnitude of the sampled prior vector $|\bar z|$ (right). }
 \label{fig:4dgenweights}
\end{figure}

\begin{figure}[htb!]
 \includegraphics[width=0.32\textwidth]{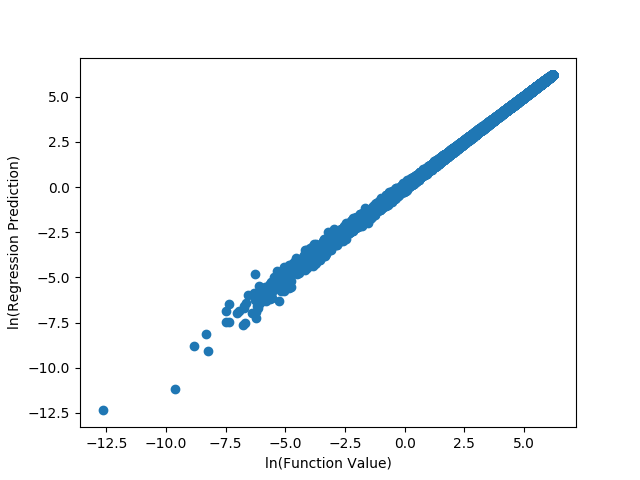}
 \includegraphics[width=0.32\textwidth]{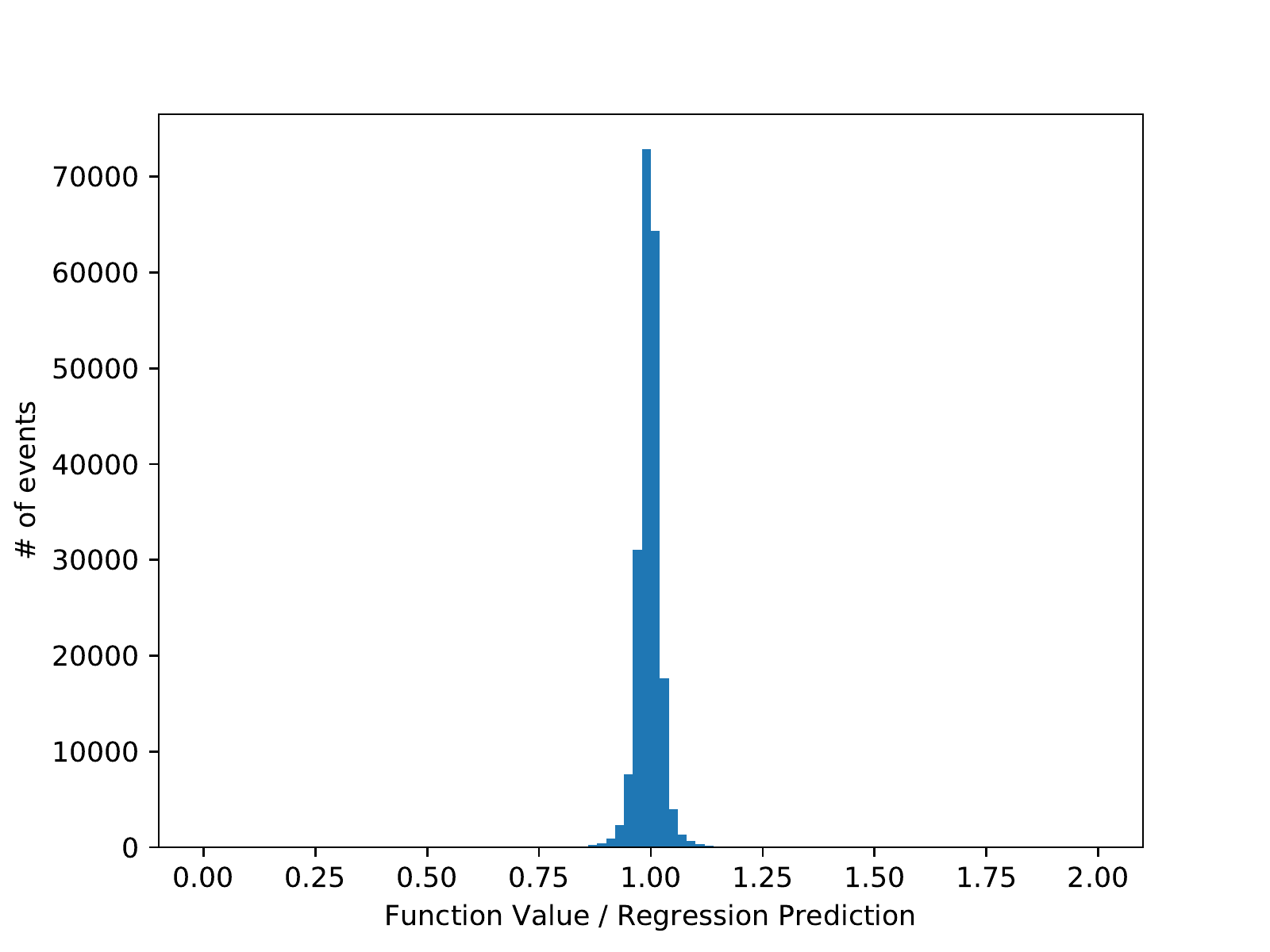}
 \includegraphics[width=0.32\textwidth]{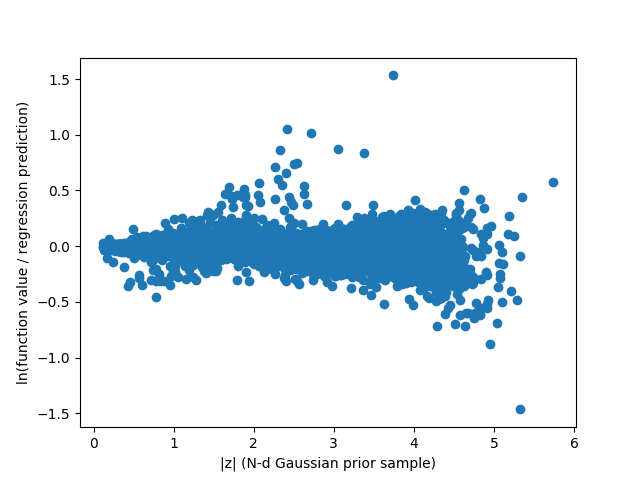}
 \caption{The correlation is shown between the regression approximation $e^{h(\bar x)}$ and the target function value $f(\bar x)$ for the 4-dimensional camel function (left).  The regression tracks the target function very well over many orders of magnitude.  The corresponding integration weight distribution is shown (center), as well as the correlation of the integration weight with the magnitude of the sampled prior vector $|\bar z|$ (right). }
 \label{fig:4dregweights}
\end{figure}

Without inverting the generative model, it is not possible to directly compute the PDF associated with the generative model along the multi-dimensional diagonal as was done for the BDT case.  Instead the correlation can be studied between the target function value and the PDF associated with the generative model as well as the regression approximation.  This correlation is shown for the generative model in Figure \ref{fig:4dgenweights} along with the integration weight distribution, as well as the correlation of the integration weight with the magnitude of the sampled prior vector $|\bar z|$.  The generating PDF tracks the target function down to relatively low values, but tends to overestimate the target in regions of very low probability.  This is in general not a major issue for Monte Carlo Integration or unweighting, since the effect on the precision and/or efficiency is small in both cases.  The correlation of the integration weights with $|\bar z|$ are striking and suggest that tails in the weight distribution might be mitigated by modifying the sampling of $z$ at integration or event generation time to modestly oversample the tails of the distribution in $|\bar z|$.  The same set of plots are shown for the corresponding regression model in Figure \ref{fig:4dregweights}, where the weight distribution is narrower, and the tendency to overestimate the target function in low probability regions is not present.

\begin{figure}[htb!]
 \includegraphics[width=0.32\textwidth]{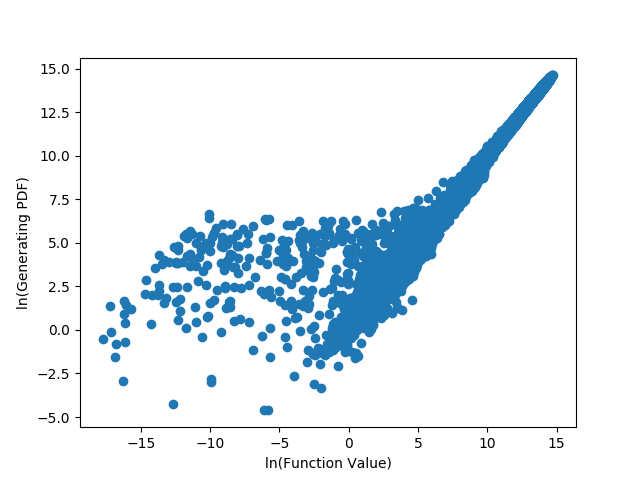}
 \includegraphics[width=0.32\textwidth]{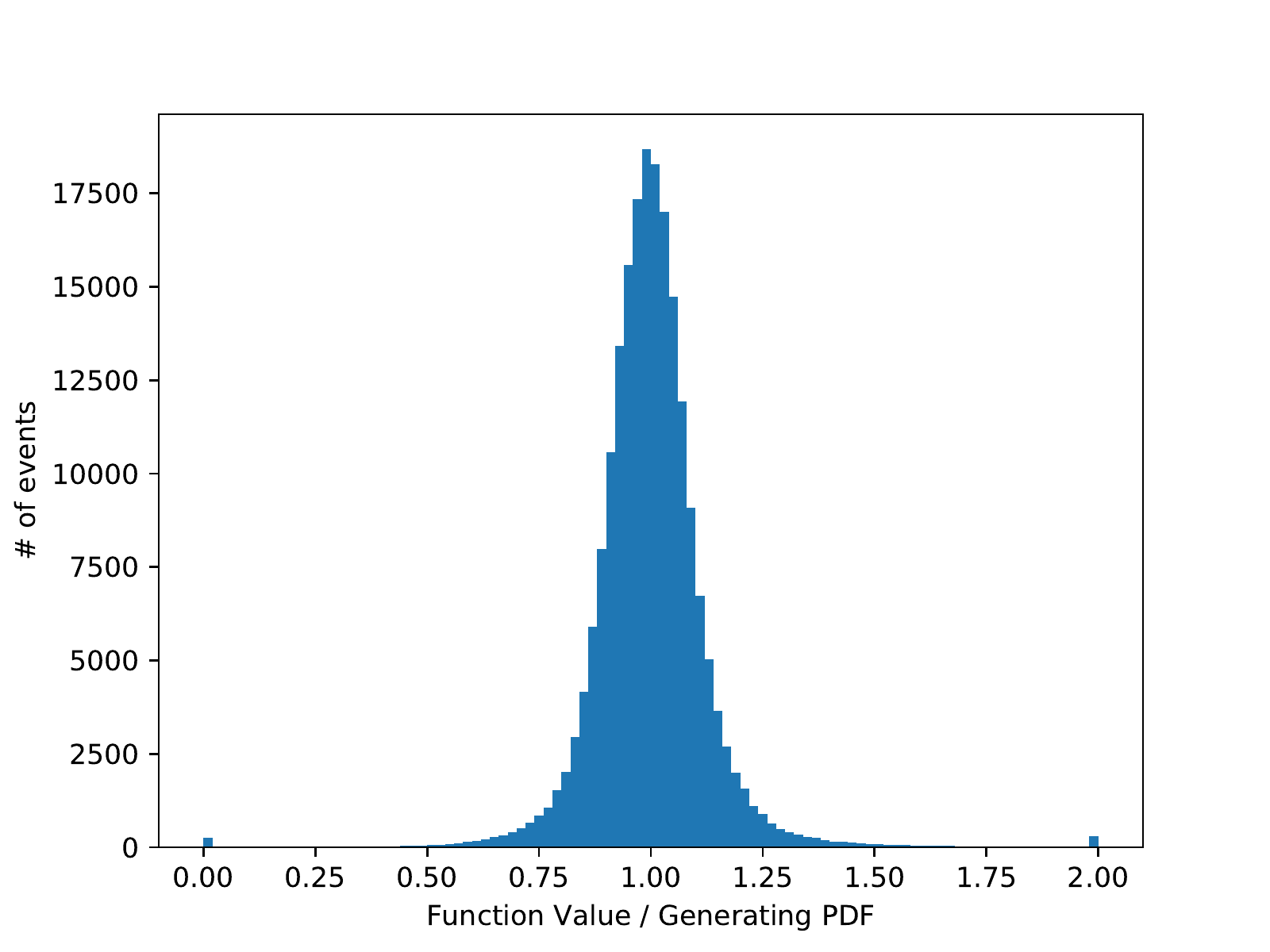}
 \includegraphics[width=0.32\textwidth]{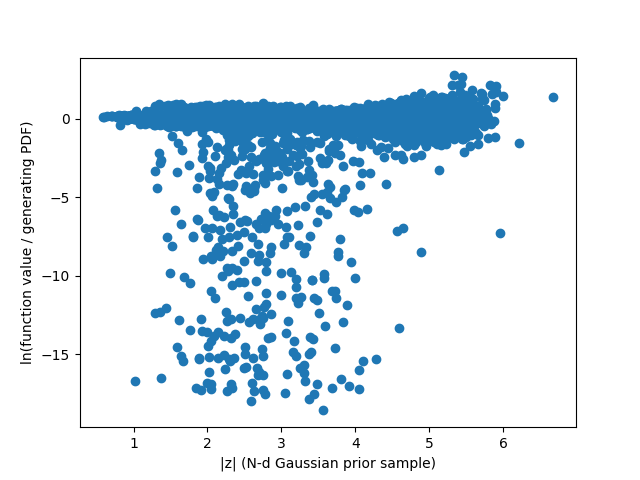}
 \caption{The correlation is shown between the PDF associated with the generative model $p_g(\bar x)$ and the target function value $f(\bar x)$ for the 9-dimensional camel function (left).  The generating PDF again tracks the target function down to relatively low values, but tends to overestimate the target in regions of very low probability.  The corresponding integration weight distribution is shown (center), as well as the correlation of the integration weight with the magnitude of the sampled prior vector $|\bar z|$ (right). }
 \label{fig:9dgenweights}
\end{figure}

\begin{figure}[htb!]
 \includegraphics[width=0.32\textwidth]{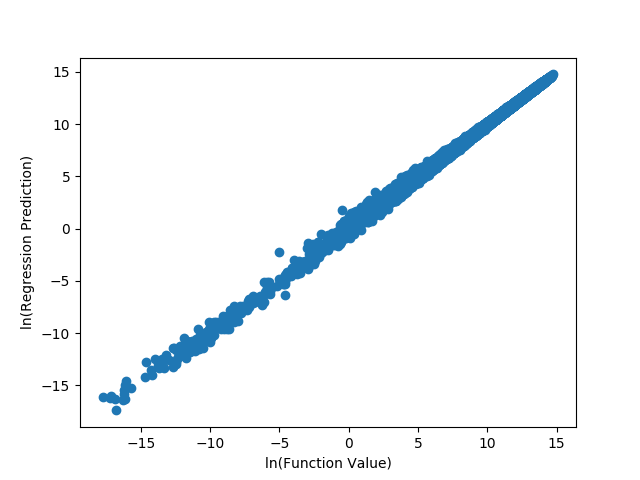}
 \includegraphics[width=0.32\textwidth]{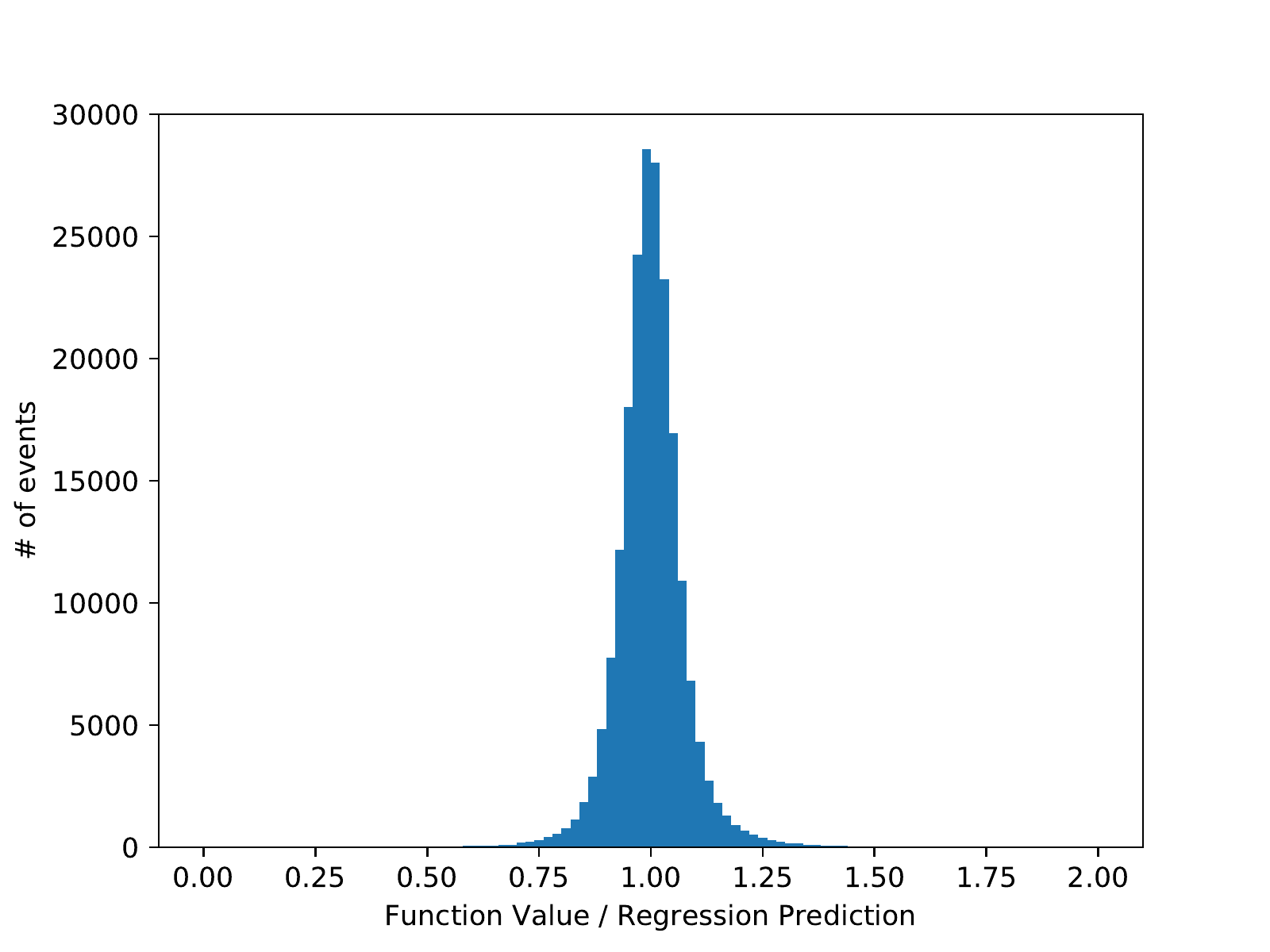}
 \includegraphics[width=0.32\textwidth]{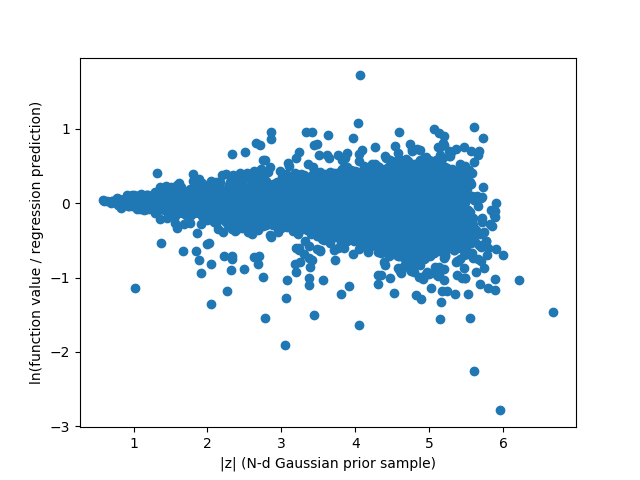}
 \caption{The correlation is shown between the regression approximation $e^{h(\bar x)}$ and the target function value $f(\bar x)$ for the 9-dimensional camel function (left).  The regression tracks the target function very well over many orders of magnitude.  The corresponding integration weight distribution is shown (center), as well as the correlation of the integration weight with the magnitude of the sampled prior vector $|\bar z|$ (right). }
 \label{fig:9dregweights}
\end{figure}

The corresponding set of plots are shown in Figures \ref{fig:9dgenweights} and \ref{fig:9dregweights} for the 9-dimensional case, where a total of 294,912 function evaluations have again been used during the training.  The qualitative behaviour is similar to the 4-dimensional case.

\section{Results and Performance Comparisons}

The performance of the new BDT and DNN-based integration algorithms are compared with VEGAS and FOAM for the multi-dimensional camel function, with the comparison carried out in terms of the number of evaluations of the target function relative to the resulting integration weight variance, which determines the achievable integration precision for a given number of additional samples.  The comparison is shown for the four-dimensional case in Table \ref{tab:camel4d} where VEGAS results are taken directly from \cite{vegas} and FOAM results from \cite{foam}.

\begin{table}[htb!]
\begin{center}
 \begin{tabular}{l|r|r|r}
 Algorithm & \# of Func. Evals & $\sigma_w/<w>$ & $\sigma_{I}/I$ \\
           &                   &                & (2e6 add. evts) \\\hline
 VEGAS & 300,000 & 2.820 & $\pm 2.0\times 10^{-3}$ \\
 Foam  & 3,855,289 & 0.319 & $\pm 2.3\times 10^{-4}$\\
 Generative BDT  & 300,000 & 0.082 & $\pm 5.8\times 10^{-5}$\\
 Generative BDT (staged) & 300,000 & 0.077 & $\pm 5.4\times 10^{-5}$\\
 Generative DNN & 294,912 & 0.083 & $\pm 5.9\times 10^{-5}$\\
 Generative DNN (staged) & 294,912 & 0.030 & $\pm 2.1\times 10^{-5}$\\
 \end{tabular}
 \caption{Performance comparison for integration of the 4-dimensional camel function between VEGAS, Foam, and the new BDT and DNN-based integration algorithms.  The comparison shows the number of function evaluations required to train or construct grids for each algorithm, the resulting integration weight variance for further generated samples after training/grid construction, and the achievable integral precision with 2 million additional samples.  The ``staged'' configurations refer to the case where integration is performed by sampling from the corresponding regression model, with the generative BDT or DNN used only to perform a secondary unweighting and integration of the regression approximation.}
 \label{tab:camel4d}
 \end{center}
\end{table}

Given the non-factorizable nature of this integrand, the performace of VEGAS saturates at relatively poor weight variance.  Foam achieves a significantly better precision, but requires significantly more function evaluations.  The machine learning based algorithms achieve a significant $\sim 4x$ improvement in precision with more than an order of magnitude reduction in the number of required function evaluations, with the generative BDT and DNN models performing similarly in this case, though the DNN regression is performing better than the BDT regression in the staged integration case.

A similar comparison is shown for the 9-dimensional camel function in Table \ref{tab:camel9d}.  The machine learning algorithms are again performing significantly better than the factorized VEGAS algorithm as expected for a high-dimensional non-factorizable integrand.  The performance of the DNN-based models is significantly better than the BDT's in this case, indicating a much better scaling with dimensionality.

\begin{table}[htb!]
\begin{center}
 \begin{tabular}{l|r|r|r}
 Algorithm & \# of Func. Evals & $\sigma_w/<w>$ & $\sigma_{I}/I$ \\
           &                   &                & (2e6 add. evts) \\\hline
 VEGAS & 1,500,000 & 19 & $\pm 1.3\times 10^{-2}$ \\
 Generative BDT  & 3,200,000 & 0.63 & $\pm 4.5\times 10^{-4}$\\
 Generative BDT (staged)  & 3,200,000 & 0.31 & $\pm 2.2\times 10^{-4}$\\
 Generative DNN  & 294,912 & 0.15 & $\pm 1.1\times 10^{-4}$\\
 Generative DNN (staged) & 294,912 & 0.081 & $\pm 5.7\times 10^{-5}$\\
 \end{tabular}
 \caption{Performance comparison for integration of the 9-dimensional camel function between VEGAS, and the new BDT and DNN-based integration algorithms.  The comparison shows the number of function evaluations required to train or construct grids for each algorithm, the resulting integration weight variance for further generated samples after training/grid construction, and the achievable integral precision with 2 million additional samples.  The ``staged'' configurations refer to the case where integration is performed by sampling from the corresponding regression model, with the generative BDT or DNN used only to perform a secondary unweighting and integration of the regression approximation.}
 \label{tab:camel9d}
 \end{center}
\end{table}
 
\section{Conclusions and Outlook}

New machine learning based algorithms have been developed and tested for Monte Carlo integration based on generative Boosted Decision Trees and Deep Neural Networks.  Both of these algorithms exhibit substantial improvements compared to existing algorithms for non-factorizable integrands in terms of the achievable integration precision for a given number of target function evaluations.  These algorithms could be extended to the application of unweighting by accept-reject sampling, where optimal performance in this regard requires an appropriate modification of the relevant loss functions in order to target unweighting efficiency rather than weight variance for integration.

There is a great deal of flexibility in terms of loss functions, network architecture and minimization for the DNN-based algorithms, where additional performance improvements could be possible beyond what has already been demonstrated in this work.  Given the increased flexibility taken together with better scaling with dimensionality, and better synergy with ongoing data science and machine learning research outside of high energy physics, the algorithms based on Deep Neural Networks are preferred over the BDT-
based algorithms, and will be the focus of future work.

Evaluating the performance on real-life examples relevant to high energy physics, namely multi-leg matrix elements for collider processes is the target of follow-up work.  It is important to understand in particular the relationship between improved performance for non-factorizable integrands, and existing transformation and multi-channeling techniques which are used to partially factorize the integration of matrix elements with VEGAS-based algorithms in existing Monte Carlo generators such as Madgraph5\_aMC@NLO\cite{mg5amc} and Sherpa\cite{sherpa}, and to what extent these types of transformations or multi-channeling may be useful in the machine learning context and how they may be incorporated.  Large scale generation of more complex processes with improved efficiency can be achieved by implementing these algorithms into commonly used matrix element Monte Carlo generators once their robustness is demonstrated and performance validated for the relevant classes of matrix elements.

\acknowledgments
This project is supported by the United States Department of Energy, Office of High Energy Physics Research
under Contract No. DE-SC0011925 and Fermi Research Alliance, LLC under Contract No. DE-AC02-07CH11359.

\bibliographystyle{JHEP}
\bibliography{mlintegrationNov2016}

\end{document}